\documentclass[12pt]{article}
\usepackage{amsfonts,colordvi}
\usepackage{amsmath,amssymb,bbm,graphicx,amstext,latexsym,xspace}
\usepackage{color}
\newcommand{\eq}{\begin{equation}}
\newcommand{\en}{\end{equation}}
\newcommand{\eqn}{\begin{eqnarray}}
\newcommand{\enn}{\end{eqnarray}}
\newcommand{\nn}{\nonumber }
\newcommand{\beq}{\begin{equation}}
\newcommand{\eeq}{\end{equation}}

\def\k0{\ensuremath{\check{0}}}

\def\cD{{\cal D}}
\def\cF{{\cal F}}

\def\cM{{\cal M}}

\def\cP{{\cal P}}
\def\cQ{{\cal Q}}

\def\c{\gamma}

\def\d{\delta}
\def\eps{\epsilon}

\def\nn{\nonumber}
\def\beq{\begin{equation}}
\def\eeq{\end{equation}}
\def\bea{\begin{eqnarray}}
\def\eea{\end{eqnarray}}
\newcommand{\w}[1]{\\[0.#1cm]}
\newcommand{\eqq}[1]{(\ref{#1})}

\def\tr{{\rm tr}}

\def\bs#1\es{\begin{split}#1\end{split}}
\def\bs#1\es{\begin{split}#1\end{split}}

\def\s2{\sqrt{2}}

\newcommand{\ft}[2]{{\textstyle\frac{#1}{#2}}}

\newcommand{\CC}{{\cal C}}
\newcommand{\KK}{{\cal K}}

\makeatletter \@addtoreset{equation}{section} \makeatother
\renewcommand{\theequation}{\thesection.\arabic{equation}}

\begin{document}


\begin{center}

\hfill ENSL-00544938 \\
\hfill MIFPA-10-55\\
\vskip .2cm
\begin{LARGE}
\textbf{On the Magical Supergravities\\[1ex] in Six Dimensions}
\end{LARGE}\\
\vspace{.8cm}
\begin{large}
M. G\"{u}naydin$^{\dagger}$\ , H. Samtleben$^{\ddagger}$ and E. Sezgin$^{*}$ \end{large}
\vspace{.35cm} \\

$^{\dagger}$ \emph{Center for Fundamental Theory \\ Institute for Gravitation and the Cosmos \\ Pennsylvania State University, University Park, PA 16802, USA\\ \rm{email:} {\tt  murat@phys.psu.edu}} \\
\vspace{.1cm}
\vspace{.3cm} $^{\ddagger}$ \emph{ Universit\'e de Lyon,
Laboratoire de Physique, UMR 5672, CNRS,\\Ecole Normale Sup\'erieure de Lyon,
F-69364 Lyon cedex 07, France,\\
Institut Universitaire de France\\
{\em email:} {\tt henning.samtleben@ens-lyon.fr}}\\
\vspace{.1cm}
\vspace{.3cm}
$^*$ \emph{George P. and Cynthia W. Mitchell Institute \\for Fundamental
Physics and Astronomy \\
Texas A\&M University, College Station, TX 77843-4242, USA\\
{\em email:} {\tt sezgin@tamu.edu} }

\vspace{.8cm} {\bf Abstract}
\end{center}
\begin{small}

Magical supergravities are a very special class of supergravity theories whose  symmetries and matter content  in various dimensions correspond  to symmetries and  underlying algebraic structures of the remarkable geometries of  the Magic Square of Freudenthal, Rozenfeld and Tits. These symmetry groups include the exceptional groups and some of their special subgroups.
In this paper, we study the general gaugings of these theories in six dimensions which lead to new couplings
between vector and tensor fields.
We show that in the absence of hypermultiplet couplings the gauge group is uniquely determined by a maximal set of commuting
translations within the isometry group $SO(n_T,1)$ of the tensor multiplet sector.
Moreover, we find that in general the gauge algebra allows for central charges that may have nontrivial action on the hypermultiplet scalars.
We determine the new minimal couplings, Yukawa couplings and the scalar potential.
\end{small}


\thispagestyle{empty}

\tableofcontents

\renewcommand{\theequation}{\arabic{section}.\arabic{equation}}

\newpage


\section{Introduction}


There exists a remarkable class of supergravity theories in $D=3,4,5,6$, known as magical supergravities, ~\cite{Gunaydin:1983rk,Gunaydin:1983bi} whose geometries and symmetries correspond to those
the Magic Square of Freudenthal, Rozenfeld and Tits \cite{MR0170974,MR0146231,MR0077076}.
In five dimensions these theories describe the coupling of $N=2$ supergravity  to 5, 8 , 14 and 26 vector multiplets, respectively,  and
 are the unique {\it unified}  Maxwell-Einstein supergravity theories with symmetric target spaces.
In $D=6$ they describe the coupling of  a fixed number of vector multiplets as well as tensor multiplets to supergravity \cite{Romans:1986er}. The scalar fields of these theories parametrize certain symmetric spaces in $D=3,4,5$ \cite{Gunaydin:1983rk} that were later referred to as very special quaternionic K\"ahler, very special K\"ahler and very special real, respectively.  Very special geometries have been studied extensively \cite{Cremmer:1984hj,deWit:1992wf,deWit:1992cr,deWit:1991nm}.
See~\cite{VanProeyen:2001wr} for a review of these geometries, their relation to $6D$ theories and a more complete list of  references on the subject.\footnote{ In this paper we are only interested in the magical supergravity theories. The conditions for oxidation of a generic real geometry to six dimensions were studied in \cite{Antoniadis:1996vw,Ferrara:1996wv}. The general conditions for oxidation of  theories with 8 supercharges and symmetric  target spaces , which include the magical theories, were studied in  \cite{Keurentjes:2002rc,Keurentjes:2002xc}. }
The magical theories in $D=6$ are parent theories from which all the magical supergravities in $D=3,4,5$ can be obtained by dimensional reduction. The scalar coset spaces in all magical supergravities are collected in table~\ref{tab:cosets}. Stringy origins and constructions of some of the magical supergravity theories in various dimensions, with or without  additional hypermultiplet couplings, are known \cite{Sen:1995ff,talkparis,Dolivet:2007sz,Bianchi:2007va,todorov-2008,Gunaydin:2009pk}.

\begin{table}[htb]
\begin{center}
\begin{tabular}{ccccccc}
$D=6$ &&$D=5$ && $D=4$ && $D=3$
\\[.6ex]\hline\\[0ex]
$\frac{SO(9,1)}{SO(9)}$ &
$\longrightarrow$ &
$\frac{E_{6(-26)}}{F_4}$ &
$\longrightarrow$ &
$\frac{E_{7(-25)}}{E_6 \times SO(2)}$ &
$\longrightarrow$ &
$\frac{E_{8(-24)}}{E_7 \times SU(2)}$
\\[2ex]
$\frac{SO(5,1)}{SO(5)}$ &
$\longrightarrow$ &
$\frac{SU^*(6)}{USp(6)}$ &
$\longrightarrow$ &
$\frac{SO^*(12)}{U(6)}$ &
$\longrightarrow$ &
$\frac{E_{7(-5)}}{SO(12) \times SU(2)}$
\\[2ex]
$\frac{SO(3,1)}{SO(3)}$ &
$\longrightarrow$ &
$\frac{SL(3,\mathbb{C})}{SO(3)}$ &
$\longrightarrow$ &
$\frac{SU(3,3)}{SU(3)\times SU(3)\times U(1)}$ &
$\longrightarrow$ &
$\frac{E_{6(+2)}}{SU(6) \times SU(2)}$
\\[2ex]
$\frac{SO(2,1)}{SO(2)}$ &
$\longrightarrow$ &
$\frac{SL(3,\mathbb{R})}{SO(3)}$ &
$\longrightarrow$ &
$\frac{Sp(6,\mathbb{R})}{U(3)}$ &
$\longrightarrow$ &
$\frac{F_{4(+4)}}{USp(6)\times USp(2)}$
\\[2ex]
\hline
\end{tabular}
\end{center}
\label{tab:cosets}
\caption{Scalar target spaces of magical supergravities in $6,5,4$ and $3$  dimensions.}
\end{table}

Gaugings of magical supergravities have been investigated in $D=5$~\cite{Gunaydin:1984ak,Gunaydin:1999zx,Gunaydin:2003yx}
as well as in $4$ and $3$ dimensions~\cite{Andrianopoli:1996cm,Gunaydin:2005bf, Gunaydin:2005df,deWit:1992up,deWit:2003ja}. However, the gaugings associated with the isometries of the scalar cosets  in $D=6$ listed above have not been studied so far. In this paper, we aim to close this gap. The gauging phenomenon is especially interesting in this case since it involves tensor as well as vector multiplets such that  the corresponding tensor and vector fields  transform in the vector and spinor representations of the isometry group $SO(n_T,1)$, respectively. Furthermore, the vector multiplets do not contain any scalar fields. Including the coupling of hypermultiplet couplings introduces additional subtleties with regard to the nature of full gauge group that are allowed by supersymmetry.

We determine the general gauging of magical supergravities in six dimensions and show that in the absence of hypermultiplet couplings the gauge group is uniquely determined by the maximal set of $(n_T-1)$ commuting translations within the isometry group $SO(n_T,1)$.
In addition, a linear combination of these generators may act on the fermion fields as a $U(1)_R$ generator of the R-symmetry group $Sp(1)_R$.
In the general case, the gauge algebra allows for central charges that may have nontrivial action on the hypermultiplet scalars.  We show that the emergence of central charges can be explained by the fact that the gauge group is a diagonal subgroup of $(n_T-1)$ translational isometries and  $(n_T-1)$ Abelian  gauge symmetries of the vector fields. 

The plan of the paper is as follows. In the next section, we give a review of the magical supergravity theories in six dimensions. In section~3 we determine the possible gauge groups and non-abelian tensor gauge transformations using the embedding tensor formalism and show that they are characterized by the choice of a constant spinor of $SO(n_T,1)$. We also give a study of the relevant spinor orbits. In section~4, we elaborate on the structure of the gauge group by embedding the symmetries of the $6D$ magical theories in the corresponding $5D$ magical supergravities. We then choose a particular basis and evaluate the gauge group generators in the vector/tensor- and hypersector. We then work out the Yukawa couplings and the scalar potential induced by the gauging. We conclude with comments on salient features of our results and open problems  as well as a discussion of the stringy origins of magical supergravity theories in section~5.

\renewcommand{\theequation}{\arabic{section}.\arabic{equation}}
\section{ Ungauged $6D$ Magical Supergravity Theories }


\subsection{Field Content of $6D$ Magical Supergravity Theories}

We consider  the minimal chiral $N=(1,0)$ supergravity in 6D coupled to $n_T$ tensor multiplets, $n_V$ vector multiplets and $n_H$ hypermultiplets \cite{Nishino:1986dc,Sagnotti:1992qw,Ferrara:1996wv,Nishino:1997ff,Ferrara:1997gh,Riccioni:2001bg}. We shall group together the single 2-form potential of pure supergravity that has self-dual field strength, with $n_T$ 2-form potentials of the tensor multiplets that have anti-selfdual field strengths, and label them collectively as $B_{\mu\nu}^I$. Thus, the field content is
\bea
\mbox{supergravity and tensor multiplets}: &&
\{ e_\mu^m, \psi_\mu^i, B_{\mu\nu}^I, \chi^{ai}, L^I\}\;,
\nonumber\\[1ex]
\mbox{vector multiplets}: &&
\{ A_\mu^A, \lambda^{Ai}\}
\;,
\nonumber\\[1ex]
\mbox{hypermultiplets}: &&
\{ \phi^X, \psi^{r}\}
\;,
\eea
with
\beq
\begin{split}
I &=0,1,\dots,n_T\ ,
\\
a &=1,\dots,n_T\ ,
\\
A&=1,\dots,n_V\ ,
\\
X&=1,\dots,4n_H\ ,
\\
r&=1,\dots,2n_H\ .
\end{split}
\eeq
The gravitino, tensorino and gaugino in addition carry the doublet index of the R-symmetry group $Sp(1)_R$ labeled by $i=1,2$. All fermions are symplectic Majorana-Weyl, where $(\psi_\mu^i, \lambda^{Ai})$ have positive chirality and $(\chi^{ai}, \psi^r)$ have negative chirality.  $L^I$ denotes a representative of the coset space $\cM_T =SO(n_T,1)/SO(n_T)$ parametrized by $n_T$ real scalars. It has the tangent space group $SO(n_T)$ with respect to which the tensorinos transform as a vector. The scalars $\phi^X$ parametrize a general quaternionic manifold~$\cM_Q$. We will discuss the structure of the scalar target spaces in the next two subsections.

Magical supergravities exist for the particular values $n_T=2, 3, 5, 9$ with the vectors transforming
in the spinor representation of $SO(n_T,1)$, see table~\ref{tab:reality} for details and their explicit
reality properties. A defining property of these theories is the existence of an $SO(n_T,1)$ invariant tensor
$\Gamma^I_{AB}$ (the Dirac $\Gamma$-matrices for $n_T=2,3 $, and Van der Waerden symbols for $n_T=5,9$, respectively), giving rise to non-trivial couplings between vector and tensor fields, and satisfying the well-known identity
\bea
\Gamma^{\vphantom{I}}_{I\,(AB}\Gamma^{I}_{C)D} &=& 0\;.
\label{magic}
\eea
These are the Fierz identities of supersymmetric Yang-Mills theories in the critical dimensions.

\begin{table}
\begin{center}
\begin{tabular}{c||c|c|c|c}
$G_T$ & ${\cal R}_{\rm v}$ & $A_\mu^A$ & $\Gamma^I_{AB}$ & ${\cal R}_{\rm ten}$ \\[1ex] \hline\hline
&&&&\\[-2ex]
${SO}(9,1)$ &
${\bf 16}_c$
& MW & $\Gamma^I_{AB}$
& ${\bf 10}$
\\[1ex]\hline&&&&\\[-2ex]
${SO}(5,1)\times {USp}(2)$ &
${\bf (4}_c,{\bf 2)}$
& SMW,\, $A=(\alpha r)$ &
$\Gamma^I_{\alpha r,\beta s} = \Gamma^I_{\alpha\beta} \epsilon_{rs}$
& ${\bf (6,1)}$
\\[1ex]\hline&&&&\\[-2ex]
${SO}(3,1)\times {U}(1)$ &
${\bf (2,1)}_+ + {\bf (1,2)}_-$
&W,\, $A=\{\alpha,\dot{\beta}\}$
&
$\left(
\begin{array}{cc}
0 & \Gamma^I_{\alpha\dot{\beta}}\\
\bar\Gamma^I_{\dot\alpha{\beta}} & 0
\end{array}
\right)$
& ${\bf (2,2)_0}$
\\[3ex]\hline&&&&\\[-2ex]
${SO}(2,1)$  &
${\bf 2}$&
M & $\Gamma^I_{AB}$
& ${\bf 3}$
\\[1ex]\hline
\end{tabular}
\caption{
{\small
The first column shows the full global symmetry groups
of the magical supergravities, the second column gives the representation
content of the vector fields under these groups, whose reality properties
are listed in the third column: Majorana (M), Weyl (W), Majorana-Weyl (MW), symplectic Majorana-Weyl (SMW).
The last column gives the two-form representation content.
 }}
\label{tab:reality}
\end{center}
\end{table}


\subsection{The Tensor Multiplet Scalars}


The $n_T$ scalars in the model parametrize the coset $\cM= SO(n_T,1)/SO(n_T)$.
It is convenient to introduce the coset representatives in the $n_T+1$ dimensional representation of the isometry group. We denote them by $(L_I, L_I^a)$ and they obey the relations~\cite{Romans:1986er}
\beq
\bs
& L^I L_I =-1\ ,\qquad L^I_a L_{Ib} = \delta_{ab}\ ,\qquad L^I L_{Ia}=0\ ,
\w2
& I = 0,1,\dots,n_T\ ,\quad  a=1,\dots,n_T\ .
\label{LLrelations}
\es
\eeq
Equation (\ref{LLrelations}) can be equivalently written  as
\bea
-L_I L_J +L_I^a L_J^a &=& \eta_{IJ}
\;,
\eea
and the coset representative can be used to define the metric
\bea
g_{IJ} &=& L_I L_J + L_I^a L_J^a
\;,
\label{defg}
\eea
and the tensors
\bea
m_{AB} &\equiv& L_I \Gamma^I_{AB} \;,\qquad
m^a_{AB}~\equiv~
 L_I^a \Gamma^I_{AB} \;,
 \label{defMM}
\eea
with $\Gamma^I_{AB}$ as given in Table~\ref{tab:reality}, which will be used to parametrize the various couplings in the action.
Note that $m^{AB}\equiv -L_I \Gamma^I{}^{AB}$ is the inverse matrix of $m_{AB}$\,. We should stress that throughout the paper raising and lowering of the $SO(n_T,1)$ indices $I,J,\cdots $ are done with the Lorentzian metric $\eta_{IJ}$,
and not the metric $g_{IJ}$, and for the $SO(n_T)$ vector indices with $\delta_{ab}$.

Next, we define the scalar current and $SO(n_T)$ composite connection as
\beq
L^{Ia} \partial_\mu L_I= P_\mu^a\ ,\qquad L^{I[a}\,\partial_\mu L_I^{b]} = Q_\mu^{\,ab}\ ,
\label{defPQ}
\eeq
where the covariant derivative in $D_\mu P_\nu^a$ involves the connection $Q_\mu^{ab}$. Integrability relations state that
\beq
D_{[\mu} P_{\nu]}^a =0\ , \qquad
Q^{\,ab}_{\mu\nu} \equiv 2\partial_{[\mu}Q_{\nu]}^{\,ab} + 2Q_{[\mu}^{\,ac} Q_{\nu]}^{\,cb}
=- 2P_{[\mu}^a P_{\nu]}^b \ .
\label{intPQ}
\eeq
It also follows from \eqq{defPQ} that
\beq
\partial_\mu L_I = P_\mu^a L_I^a\ ,\qquad D_\mu L_I^a = P_\mu^a L_I\ .
\eeq

A parametrization of the coset representative
which is convenient for the following can be given according to the
decomposition (3-grading)
\bea
\mathfrak{so}(n_T,1)&\longrightarrow&
N^-_{(n_{T}-1)} \oplus \left(\mathfrak{so}(n_{T}-1)\oplus \mathfrak{so}(1,1)\right) \oplus N^+_{(n_{T}-1)}
\;,
\label{Ntranslations}
\eea
where the $\pm$ superscript refers to the $\mathfrak{so}(1,1)$ charges.
We choose
\bea
L &=& e^{\varphi^\alpha\,N_\alpha}\,e^{\sigma \Delta}
\;,
\label{Lpar}
\eea
with the $(n_{T}-1)$ nilpotent generators $N_\alpha \in N^+_{(n_{T}-1)}$,
and the $\mathfrak{so}(1,1)$ generator $\Delta$, normalized such that
$[\Delta,N_\alpha]=N_\alpha$\,.
With this parametrization, we obtain
\bea
P_\mu^\alpha &=& e^{-\sigma}\,\partial_\mu \varphi^\alpha
\;,\qquad
P_\mu^{1} ~=~ \partial_\mu \sigma
\;,\qquad
Q_\mu^{1\alpha} ~=~ e^{-\sigma}\,\partial_\mu \varphi^\alpha
\;,
\label{PQexp}
\eea
where the index $a$ has been split into $a\rightarrow\{1,\alpha\}$\,,
with $\alpha=2, \dots, n_T$\,.


\subsection{ Hypermultiplet scalars }


Supersymmetry requires the hyperscalar manifold ${\cal M}_Q$ to be quaternionic K\"ahler~\cite{Bagger:1983tt}.
Let us review the basic properties of quaternionic K\"ahler manifolds, following \cite{Galicki:1985qv}.  They have the tangent space group $Sp(n_H)\times Sp(1)_R$, and one can introduce the vielbeins $V_X^{ri}$ and their inverse $V^X_{ri}$ satisfying
\bea
g_{XY} V^X_{ri} V^Y_{sj} =\Omega_{rs}\eps_{ij}\ ,
\qquad
V^X_{ri} V^{Y rj} + \ ( X \leftrightarrow Y ) = g^{XY} \d_i^j\ ,
\eea
where $g_{XY}$ is the target space metric.
An $Sp(n_H)\times Sp(1)_R$ valued connection is defined through the vanishing torsion condition~\footnote{$Sp(n)$ refers to the compact symplectic group of rank $n$ which is  denoted as $USp(2n)$ in some of the physics literature.}
\bea
\partial_X V_{Y ri} + A_{Xr}{}^s V_{Ysi} +A_{X i}{}^j V_{Y rj}\
-( X \leftrightarrow Y) =0\ .
\eea
From the fact that the vielbein $V^X_{ri}$ is covariantly constant, one derives that\footnote{
In our conventions, $[\nabla_X,\nabla_Y]\,X_Z = R_{XYZ}{}^T\,X_T$\,.}
\bea
R_{XYZT} V^T_{ri} V^Z_{sj} &=& \epsilon_{ij}\,F_{XY}{}_{rs}+\Omega_{rs}\,F_{XY}{}_{ij}
\;,
\label{Riemann}
\eea
where $F_{ij}$ and $F_{rs}$ are the curvature two-forms of the $Sp(1)_R$ and $Sp(n_H)$ connection, respectively.

The manifold has a quaternionic K\"ahler structure characterized by three locally
defined $(1,1)$ tensors $J^x{}_X{}^Y$ $(x,y,z=1,2,3)$ satisfying the quaternion algebra
\bea
J^{x}{}_X{}^Y J^{y}{}_Y{}^Z = -\delta^{xy}\delta_X^Z +\eps^{xyz} J^z{}_X{}^Z\ .
\eea
In terms of the vielbein, these tensors can be expressed as
\bea
J^x{}_X{}^Y &=& -i (\sigma^x)_i{}^j\,V_X^{ri} V^Y_{rj}
\;,
\label{cs}
\eea
with Pauli matrices $\sigma^x$\,.
We can define a triplet of two-forms $J^x_{XY}=J^x{}_X{}^Z g_{ZY}$, and these are covariantly constant as follows
\bea
\nabla_X J^x_{YZ} +\eps^{xyz} A_X^y J^z_{YZ}=0\ ,
\label{cc}
\eea
with $A_X^x \equiv \frac{i}2  (\sigma^x)_i{}^j A_{X j}{}^i$\,.
For $n_H >1$, quaternionic K\"ahler manifolds are Einstein spaces,
i.e.\ $R_{XY}=\lambda g_{XY}$. It follows, using (\ref{cc}), that \cite{Galicki:1985qv}
\bea
F^x_{XY} = \frac{\lambda}{n_H+1} J^x_{XY}\ .
\label{ks}
\eea
Local supersymmetry relates $\lambda$ to the gravitational coupling constant (which we will set to one), and in particular requires that $\lambda <0$ \cite{Bagger:1983tt}, explicitly $\lambda=-(n_H+1)$. For $n_H=1$ all Riemannian 4-manifolds are quaternionic K\"ahler. Sometimes (\ref{ks}) is used to extend the definition of quaternionic K\"ahler to 4D, which restricts the manifold to be Einstein and self-dual \cite{Galicki:1985qv}.

Homogeneous quaternionic K\"ahler manifolds were classified by Wolf \cite{Wolf:1965} and Alekseevski \cite{Alekseevskii:1975}. For $\lambda>0$, they are the well known compact symmetric spaces, and for $\lambda<0$ they are noncompact analogs of these symmetric spaces, and non-symmetric spaces found by Alekseevskii \cite{Alekseevskii:1975}. There exists an infinite family of homogeneous quaternionic K\"ahler spaces that are not in Alekseevskii's classification.  As was shown in \cite{deWit:1991nm} this infinite family of quaternionic K\"ahler spaces arises as the scalar manifolds of $3D$ supergravity theories obtained by dimensionally reducing the generic non-Jordan family of $5D$ , $N=2$ Maxwell-Einstein supergravities discovered in \cite{Gunaydin:1986fg}.

Choosing $\lambda=-(n_H+1)$, and using $M_i{}^j= -i(\sigma^x)_i{}^j M^x$ for any triplet $M^x$,
we have the relation
\beq
F_{XY i}{}^j= -2V_{[X}^{jr} V_{Y]ir}\ .
\eeq
Substitution of this relation into \eqq{Riemann} and use of curvature cyclic identity gives \cite{Bagger:1983tt}
\beq
F_{XYrs} = V_{[X}^{pi} V^q_{Y]i} \left( -2\Omega_{pr}\Omega_{qs} +\Omega_{pqrs}\right) \ ,
\eeq
where $\Omega_{pqrs}$ is a totally symmetric tensor defined by this equation.

For any isometry on the quaternionic K\"ahler manifold defined by a Killing vector field~$K^X$,
one can define the triplet of moment maps \cite{Galicki:1986ja}
\bea
C^x &\equiv& \frac1{4n_H} J^x{}_Y{}^X \nabla_X K^Y
\;,
\label{defC2}
\eea
satisfying
\bea
D_X C^x &\equiv&\partial_X C^x +\eps^{xyz} A_X^y C^z  ~=~    J^x{}_{XY}\, K^Y
\;,
\label{DCJ}
\eea
where in particular we have used (\ref{Riemann}). Using  \eqq{cs}, we can write $C_i{}^j= -i(\sigma^x)_i{}^j C^x$ as
\beq
C_i{}^j = -\frac{1}{2n_H}\, V^X_{ri} V_Y^{rj} \nabla_X K^Y\ .
\label{defC}
\eeq
As usual, these functions will later parametrize the Yukawa couplings and the scalar potential of the gauged theory.
For later use, let us also define the function
\bea
C_r{}^s &\equiv& -\ft12 V^X_{ri} V_Y^{si} \nabla_X K^Y \;,
\label{Crs}
\eea
for a given Killing vector field $K^Y$, which satisfies $\nabla_X C_r{}^s = - F_{XY r}{}^s K^Y$,
which may be shown in analogy with (\ref{DCJ}).

We conclude this section by defining the notation
\bea
P_\mu^{ri} &=& \partial_\mu \phi^X V_X^{ri}\ ,\qquad
Q_{\mu}^{ij} ~=~ \partial_\mu \phi^X A^{ij}_X \;,\qquad
Q_{\mu}^{rs} ~=~ \partial_\mu \phi^X A^{rs}_X \;,
\label{PQQhyper}
\eea
which will be used in the following sections together with the relations
\beq
D_{[\mu} P_{\nu]}^{ri} = 0\ ,\quad
Q_{\mu\nu i}{}^j = 2P_{[\mu}^{rj}P_{\nu]ri}\ ,\quad
Q_{\mu\nu rs} = P_{[\mu}^{pi} P^q_{\nu]i} \left( -2\Omega_{pr}\Omega_{qs} +\Omega_{pqrs}\right)\ .
\eeq
%


\subsection{The Field Equation and Supersymmetry Transformations}


The bosonic field equation of the full theory including the hypermultiplets are given
up to fermionic contributions by~\cite{Nishino:1997ff,Riccioni:2001bg}
\bea
0 &=& G_{\mu\nu\rho}^+ \ ,
\label{b1}
\w2
0 &=& G^{a-}_{\mu\nu\rho}\ ,
\label{b2}
\w2
0 &=& R_{\mu\nu} -\frac14 g_{I J} G_{\mu\rho\sigma} {}^I G_\nu{}^{\rho\sigma J}\, -P_\mu^a P_{\nu a} -2P_\mu^{ri} P_{\nu ri}
\nn\w2
&&
-2 m_{AB} \left(F^A_{\mu\rho} F_{\nu}{}^{\rho B} -\frac18 g_{\mu\nu}
F^A_{\rho\sigma} F^{\rho\sigma B} \right) \ ,
\label{b3}
\w2
0 &= & D_\mu P^{\mu a}
-\frac12m^a_{AB} F^A_{\mu\nu} F^{\mu\nu B} -\frac{1}{6} G_{\mu\nu\rho} ^a G^{\mu\nu\rho}\ ,
\label{b4}
\w2
0 &=& D_\mu P^{\mu ri}\ ,
\label{b5}
\w2
0 &=&
D_\nu \left(m_{AB} F^{\mu\nu B}\right)
+ \left( m_{AB} G^{\mu\nu\rho}_-+ m^a_{AB} G^{\mu\nu\rho\,a}_+ \right) F_{\nu\rho}^B \ ,
\label{b6}
\eea
where we have defined the $3$-form and $2$-form field strengths
\beq
\begin{split}
G_{\mu\nu\rho}^I &=  3 \partial_{[\mu} B^I_{\nu\rho]}
+ 3 \Gamma^I_{AB}  F^A_{[\mu\nu} A^B_{\rho]}\ ,
\w2
F^A_{\mu\nu} &=2 \partial_{[\mu} A^A_{\nu]}\ .
\label{f23}
\end{split}
\eeq
and the projected field strengths
\beq
G_{\mu\nu\rho} = G_{\mu\nu\rho}^I L_I\ ,\qquad G_{\mu\nu\rho}^a=G_{\mu\nu\rho}^I L_I^a\ ,
\eeq
and the superscripts $\pm$ in (\ref{b1}), (\ref{b2}) refer to the (anti-)selfdual part of the projected field strengths.
The covariant derivatives acting on objects carrying the tangent space indices of the tensor and hyperscalar manifolds are defined as
\bea
D_\mu X^a &=& \partial_\mu X^a +Q_\mu^{ab} X_b\ ,
\nn\w2\
D_\mu X^{ri} &=& \partial_\mu X^{ri} + Q_\mu^{rs} X_s^i + Q_\mu^{ij} X^r_j\ ,
\eea
with the connections from (\ref{defPQ}), (\ref{PQQhyper}).
The fermionic field equations, to linear order in fermionic fields, take the form
\bea
0 &=& \c^{\mu\nu\rho} D_\nu \psi_\rho^i -\frac12 G^{\mu\nu\rho} \c_\nu \psi_\rho^i -\frac12 \c^\nu\c^\mu \chi^{ai} P_\mu^{a}
+ \c^\nu \c^\mu \psi_r  P_\mu^{ri}
\nn\w2
&&
 + \frac12 m_{AB} \left( \c^{\rho\sigma} \c^\mu \lambda^{Ai}  F_{\rho\sigma}^B \right)
+ \frac14 H_{\mu\nu\rho}^a\c^{\nu\rho} \chi^{ai}\ ,
\label{f1}\w2
0 &=& \c^\mu D_\mu \chi^a - \frac{1}{24} \c^{\mu\nu\rho}
\chi^a  G_{\mu\nu\rho} -\frac12 m^a_{AB}
\c^{\mu\nu} \lambda^A F_{\mu\nu}^B \nn
\w2
&&
+\frac14 G_{\mu\nu\rho}^a \c_{\mu\nu}\psi_\rho{}
-\frac12 \c^\mu\c^\nu\psi_\mu  P_\nu^a\ ,
\label{f2}\w2
0 &=& \c^\mu D_\mu \psi^r + \frac1{24} \c^{\mu\nu\rho} \psi^r G_{\mu\nu\rho}
-\c^\mu \c^\nu \psi_{\mu i} P_\mu^{ri}\ ,
\label{f3}\w2
0 &=& m_{AB} \c^\mu D_\mu \lambda^B + \frac14  m_{aAB}
\c^{\mu\nu} \chi^a F_{\mu\nu}^B
+ \frac{1}{24} m_{aAB} \c^{\mu\nu\rho} \lambda^B G_{\mu\nu\rho}^a
\nn\w2
&& + \frac12 m_{aAB} \c^\mu \lambda^B P_\mu^a
+\frac14 m_{AB} \c^\mu \c^{\nu\rho} \psi_\mu F_{\nu\rho}^B\ ,
\label{f4}
\eea
where we have suppressed the $Sp(1)_R$ indices.
The supersymmetry transformation rules, up to cubic fermion terms, are
\beq
\begin{split}
\delta e_\mu^m  &= \bar\eps\c^m\psi_\mu\ ,
\w2
\delta \psi_\mu &= D_\mu \eps + \frac{1}{48}
\c^{\rho\sigma\tau} \c_\mu  \eps G_{\rho\sigma\tau}\ ,
\w2
\delta_{\rm cov.} B_{\mu\nu}{}^I &= - 2 \bar\eps \c_{[\mu} \psi_{\nu]}\, L^I + \bar\eps\c_{\mu\nu} \chi^a\, L^I_a\ ,
\w2
\delta \chi^a &= \frac12 \c^\mu \eps P_\mu^a - \frac{1}{24} \c^{\mu\nu\rho} \eps
G^a_{\mu\nu\rho}\ ,
\w2
\delta L_I &= \bar\eps\chi^a L_I^a\ ,
\w2
\delta A_\mu^A  &= \bar\eps \c_\mu \lambda^A \ ,
\w2
\delta\lambda^A &= - \frac14 \c^{\mu\nu} \eps F^A_{\mu\nu} \ ,
\w2
\delta\phi^X &= V^X_{ri}\,\bar{\epsilon}^{i} \psi^r\ ,
\w2
\delta\psi^r &= P_\mu^{ri} \gamma^\mu\epsilon_i\ .
\label{t7}
\end{split}
\eeq
The covariant derivative of the supersymmetry parameter carries the  Lorentz algebra valued spin connection
and the $Sp(1)_R$ connection, and the covariant variation of the 2-form potential is defined as
\beq
\delta_{\rm cov.} B_{\mu\nu}^I = \delta B_{\mu\nu}{}^I - 2 \Gamma^I_{AB} A^A_{[\mu} \delta A^B_{\nu]}\ ,
\label{deltacov}
\eeq
such that we have the general variation formula
\beq
\delta G_{\mu\nu\rho}^I = 3\partial^{\vphantom{I}}_{[\mu} \delta_{\rm cov.} B_{\nu\rho]}^I + 6 \Gamma^I_{AB} F_{[\mu\nu}^A \delta A_{\rho]}^B\ .
\eeq
The field strengths are invariant under the gauge transformations
\bea
\delta A_\mu^A &=& \partial_\mu \Lambda^A\ ,
\nn\w2
\delta_{\rm cov.} B_{\mu\nu}^I &=& 2\partial_{[\mu} \Lambda_{\nu]}^I -2\Gamma^I_{AB} \Lambda^A F_{\mu\nu}^B\ .
\label{gt_abelian}
\eea
%


\subsection{The Action}


The field equations described above are derivable from the following action \cite{Ferrara:1997gh,Riccioni:1999xq}
\footnote{In our conventions, the Minkowski metric is given by $\eta_{mn} = \hbox{diag.} (-,+,+,+,+,+)$, and the Clifford algebra is generated by $\{ \c_m, \c_n\} = 2\eta_{mn}$. The
Ricci tensor is defined as $R_\mu{}^m= R_{\mu\nu}{}^{mn}\, e^\nu_n$.
The $Sp(1)_R$ indices are raised and lowered as $\lambda^i=\epsilon^{ij}\lambda_j$, $\lambda_j=\lambda^i\epsilon_{ij}$,
with $\epsilon_{ij}\epsilon^{ik}=\delta_j^k$, and the $SO(n_T)$ indices are raised and lowered with $\delta_{ab}$. Often we will suppress the $Sp(1)_R$ indices, and use the notation $\bar\psi \chi = \bar\psi^i\chi_i$. The fermionic bilinears have the symmetry ${\bar\psi}\gamma_{\mu_1....\mu_n} \chi= (-1)^n {\bar\chi}\gamma_{\mu_n...\mu_1}\psi$, with the $Sp(1)_R$ index contraction suppressed.}
\beq
\begin{split}
e^{-1} {\cal L} &= R -\frac{1}{12}  g_{IJ} G_{\mu\nu\rho}^I G^{\mu\nu\rho J} -\frac14 P_\mu^a P^{\mu a}
-\frac12 P_\mu^{ri} P_{\mu ri}
\w2
&
-\frac14 m_{AB} F_{\mu\nu}^A F^{\mu\nu B} -\frac{1}{8}\varepsilon^{\mu\nu\rho\sigma\lambda\tau} \Gamma_{IAB} B_{\mu\nu}^I F_{\rho\sigma}^A F_{\lambda\tau}^B
\w2
& +\frac12 {\bar\psi}_\mu\gamma^{\mu\nu\rho} D_\nu\psi_\rho
 -\frac12 {\bar\chi}^a\gamma^\mu D_\mu \chi^a-\frac12 \bar{\psi}^r \gamma^\mu D_\mu \psi_r
\w2
&
- m_{AB} {\bar\lambda}^A \gamma^\mu D_\mu \lambda^B +\frac12{\bar\psi}_\mu\gamma_\nu\gamma^\mu\chi^a P^\nu_a
-(\bar{\psi}_\mu^i \gamma_\nu\gamma^\mu \psi^r) P^\nu_{ri}
\w2
& +\frac{1}{48} G_{\mu\nu\rho} \left( {\bar\psi}^\lambda\gamma_{[\lambda}\gamma^{\mu\nu\rho}\gamma_{\tau]}\psi^\tau
+{\bar\chi}^a\gamma^{\mu\nu\rho}\chi^a-{\bar\psi}^r\gamma^{\mu\nu\rho}\psi_r
\right)
\w2
& +\frac{1}{24} G_{\mu\nu\rho}^a\left(
\bar{\psi}_\lambda\gamma^{\mu\nu\rho}\gamma^\lambda\chi^a
- m_{a AB} \bar\lambda^A \gamma^{\mu\nu\rho} \lambda^B\right)
\w2
&-\frac12 F_{\mu\nu}^A \left( m_{AB}\,{\bar\psi}_\lambda \gamma^{\mu\nu}\gamma^\lambda \lambda^B - m_{a AB}\, {\bar\chi}^a \gamma^{\mu\nu} \lambda^B\right)\ ,
\label{action}
\end{split}
\eeq
provided that the (anti-)selfduality conditions \eqq{b1}, \eqq{b2} are imposed \emph{after} the variation of the action with respect to the $2$-form potential. In particular, the 2-form field equation, upon projections with $L_I$ and $L_I^a$ yields
\bea
&& \nabla_\mu G^{\mu\rho\sigma} +P_\mu^a G_a^{\mu\rho\sigma} + \frac14 \varepsilon^{\rho\sigma\lambda\tau\mu\nu} m_{AB} F_{\lambda\tau}^A F_{\mu\nu}^B=0\ ,
\nn\w2
&& \nabla_\mu G^{\mu\rho\sigma\,a} + P_\mu^a G^{\mu\rho\sigma}  -\frac14 \varepsilon^{\rho\sigma\lambda\tau\mu\nu} m^a_{AB} F_{\lambda\tau}^A F_{\mu\nu}^B=0\ ,
\eea
up to fermionic contributions.
These equations, in turn, agree with the results that follow from taking the divergence of the (anti-)selfduality equations \eqq{b1} and \eqq{b2}.

The presence of the $B\wedge F\wedge F$ term in the action is noteworthy. Since the $3$-form field strength is Chern-Simons modified, normally it is not expected to arise in the action because in this case the 2-form potential transforms under Yang-Mills gauge transformations which typically do not leave invariant a term of the form $B\wedge F\wedge F$ in the action. However, this term is allowed in magical  supergravities due to the identity \eqq{magic}.


\section{Gauging a Subgroup of the Global Symmetry Group}


We begin with the building blocks
needed for the gauging of a subgroup $G_0$ of the global symmetry  group of the Lagrangian that utilizes a suitable subset of the $n_V$ vector fields that is dictated by the so called embedding tensor~\cite{deWit:2002vt,deWit:2004nw,deWit:2005hv}, which is subject to certain constraints.
The global symmetry  group of the Lagrangian (\ref{action}) and hence the equations of motion that follow from it obviously contains the isometry group ${SO}(n_T,1)$ of the tensor scalars. For the magical theories with  $n_T=5$ and $n_T=3$  it comprises an additional factor ${USp}(2)$ and ${U}(1)$, respectively, exclusively acting on the vector multiplets. In addition, all these theories have an $Sp(1)_R$ R-symmetry group and $U(1)^{n_V}$ Abelian  symmetry groups.

Most of the formulas presented are very similar to the structures encountered in the gauging of the maximal supergravity in six
dimensions~\cite{Bergshoeff:2007ef}, we shall see however that in contrast to the maximal case,
the construction for the magical theories allows only for a very limited choice of possible gauge groups.

\subsection{Embedding Tensor and the Tensor Hierarchy}

The key ingredient in the construction is the general covariant derivative
\beq
\cD_{\mu}= \partial_{\mu}-A_\mu{}^A\,X_A\ ,
\label{et}
\eeq
where
\bea
X_A = \Theta_A{}^{IJ}\,t_{IJ}+\Theta_A{}^{\cal X}  t_{\cal X}+\Theta_A{}^{\cal A}  t_{\cal A}
\ ,
\label{generator}
\eea
showing that the gauge group is parametrized by the choice of
the embedding tensors $\Theta_A{}^{IJ}$, $\Theta_A{}^{\cal X}$, and $\Theta_A{}^{\cal A}$.
Here, $t_{IJ}=t_{[IJ]}$ are the ${SO}(n_T,1)$ generators satisfying the algebra
\bea
{}[\,t_{IJ},t_{KL}]&=& 4\,(\eta_{I[K}\,t_{L]J}-\eta_{J[K}\,t_{L]I})\ ,
\eea
while the generators $t_{\cal X}$ span the additional symmetries ${USp}(2)$ and ${U}(1)$ for
$n_T=5$ and $n_T=3$, respectively.
The generators $t_{\cal A}$ denote the isometries of the quaternionic K\"ahler manifold parametrized by the hyperscalars, including the $Sp(1)_R$ R-symmetry. We will denote the group with generators $(t_{IJ}, t_{\cal X})$ by $G_T$ (see Table 1) and the group with generators  $t_{\cal A}$ by $G_H$.

For transparency of the presentation we will first discuss the case of gauge groups that do not involve the hyperscalars, i.e.\ set $\Theta_A{}^{\cal A}=0$, and extend the construction
to the general case with $\Theta_A{}^{\cal A}\not=0$ in section~\ref{sec:gaugehypers}.

Closure of the gauge algebra imposes the conditions~\cite{deWit:2005hv,deWit:2008ta}
\bea
[X_A,X_B\,] &=& -X_{[AB]}{}^C \,X_C\ ,\qquad X_{(AB)}{}^C X_C ~=~ 0
\;,
\label{quadconstraint}
\eea
where the ``structure constants''
$X_{AB}{}^C \equiv (X_A)_B{}^C$ are obtained from
the generator (\ref{generator}) evaluated in the
 representation ${\cal R}_v$ of the vector fields and are in general not antisymmetric in $A$ and $B$.
The proper non-abelian field strength
transforming covariantly under gauge transformations is given by the combination~\cite{deWit:2004nw,deWit:2008ta}
\beq
{\cal G}_{\mu\nu}^A =  {\cal F}_{\mu\nu}^A + X_{(BC)}{}^A\,B^{BC}_{\mu\nu}\ ,
\label{defG}
\eeq
where
\beq
{\cal F}_{\mu\nu}^A =  2\partial_{[\mu} A_{\nu]}^A + X_{[BC]}{}^A A_\mu^B A_\nu^C\ .
\eeq
The two-forms $B^{AB}_{\mu\nu}=B^{(AB)}_{\mu\nu}$ transform in the symmetric tensor product
of two vector representations $({\cal R}_{\rm v}\otimes {\cal R}_{\rm v})_{\rm sym}$,
and the non-abelian gauge transformations are
\bea
\delta A_\mu^A &=& {\cal D}_\mu \Lambda^A  - X_{(BC)}{}^A\, \Lambda_\mu^{(BC)}
\;,\nonumber\\
\delta B_{\mu\nu}^{AB} &=& 2{\cal D}_{[\mu}\Lambda_{\nu]}^{AB}
-2\Lambda^{(A} {\cal G}_{\mu\nu}^{B)}
+2 A_{[\mu}^{(A} \delta A_{\nu]}^{B)}
\;.
\label{fullgt}
\eea
Consistency of the construction imposes that the additional two-forms $B^{(AB)}_{\mu\nu}$ required in
(\ref{defG}) for closure of the non-abelian gauge algebra on the vector fields
form a subset of the $n_T$ two-forms present in the theory. In other words, it is necessary
that the intertwining tensor $X_{(BC)}{}^A$ factors according to
\bea
X_{(BC)}{}^A &=&\Gamma^I_{BC}\,\theta_I^A
\;,
\label{lincon1}
\eea
with a constant tensor $\theta_I^A$ such that with the identification
$B^I_{\mu\nu}=\Gamma^I_{AB} B_{\mu\nu}^{AB}$ the system of gauge transformations (\ref{fullgt}) takes the form
\bea
\delta_\Lambda A_\mu^A &=& {\cal D}_\mu \Lambda^A  - \theta_I^A \Lambda_\mu^I
\;,\nonumber\\
\delta_{{\rm cov},\,\Lambda} B_{\mu\nu}^{I} &=& 2{\cal D}_{[\mu}\Lambda_\nu^{I}
-2\Gamma^I_{AB}\Lambda^{A} {\cal G}_{\mu\nu}^{B}
\;,
\eea
with $\delta_{\rm cov}$ defined in (\ref{deltacov}),
and provides a proper covariantization of the {\it Abelian} system (\ref{gt_abelian}).
This shows how the gauging of the theory in general not only corresponds to covariantizing the derivatives according to (\ref{et}) but also induces a nontrivial deformation of the $2$-form tensor gauge transformations. In particular, $2$-forms start to transform by (St\"uckelberg)-shift under the gauge transformations of the $1$-forms. Pushing the same reasoning to the three-form potential and the associated gauge transformations, leads to the following set of covariant field strengths

\bea
{\cal G}_{\mu\nu}^{A} &\equiv & 2\partial_{[\mu} A_{\nu]}^A + X_{[BC]}{}^A A_\mu^B A_\nu^C
+ B_{\mu\nu}^I \theta_I^A\ ,
\nn\w2
{\cal H}^I_{\mu\nu\rho} &\equiv&
3\, {\cal D}_{[\mu} B^I_{\nu\rho]} +
6\,\Gamma^I_{AB}\,A_{[\mu}{}^A\,\Big(\partial_{\nu} A_{\rho]}{}^B+
\ft13  X_{[CD]}{}^B A_{\nu}{}^C A_{\rho]}{}^D\Big)
+\,\theta^{IA}\, C_{\mu\nu\rho\,A} \ ,
\nn\w2
{\cal G}_{\mu\nu\rho\sigma\,A} &\equiv&
4{\cal D}_{[\mu} C_{\nu\rho\sigma]\,A}
-(\Gamma_I)_{AB} \Big( 6 B^I_{\mu\nu} {\cal G}_{\rho\sigma}^B+6 \theta^{BJ} B^I_{[\mu\nu}
B^{\vphantom{I}}_{\rho\sigma] J}
\nn\w2
&& {}+8\Gamma^I_{CD} A_{[\mu}^B A_\nu^C \partial_\rho A_{\sigma]}^D +2 \Gamma^I_{CD} X_{EF}{}^{D} A_{[\mu}{}^B A_\nu{}^C A_\rho{}^E A_{\sigma]}{}^F\Big)\ ,
\label{covAH}
\eea
with three-form fields $C_{\mu\nu\rho\,A}$.
While the construction so far is entirely off-shell, the equations of motion will impose (anti-)self-duality of the
dressed field strengths $L_I{\cal G}^I_{\mu\nu\rho}$ and $L^a_I{\cal G}^I_{\mu\nu\rho}$, respectively,
whereas the three-form fields $C_{\mu\nu\rho\,A}$ are on-shell dual to the vector fields by means of a first order equation
\bea
e\,\theta_I^A{\cal G}^{\mu\nu\rho\lambda}_{A} &=& -\frac12\, \,\epsilon^{\mu\nu\rho\lambda\sigma\tau}\,m_{AB}\,{\cal G}_{\sigma\tau}^A \theta_I^B\ ,
\label{de}
\eea
with the metric $m_{AB}$ from (\ref{defMM}).
In particular, the three-form fields  transform in the contragredient representation under the global symmetry group. Similar to the above construction, their presence in the first equation of (\ref{covAH}) is required for closure of the algebra on the two-forms.\footnote{Strictly speaking, also the four-form field strength ${\cal F}_{\mu\nu\rho\sigma\,A}$ needs to be corrected by a St\"uckelberg type term carrying explicit four forms that are on-shell duals to the scalar fields. For our present purpose we will ignore these terms as they are projected out from all the relevant equations of motion.}
The hierarchy of $p$-forms may be continued to four-forms and five-forms
which are on-shell dual to the scalar fields and the embedding tensor, respectively, see~\cite{deWit:2008ta,Hartong:2009vc}, but none of these fields will enter the covariantized action and the tensor hierarchy can consistently be truncated to (\ref{covAH}).

The field strengths \eqq{defG} and (\ref{covAH}) transform covariantly under the full set of non-abelian gauge transformations
\bea
\delta_\Lambda A_\mu^A &=&
{\cal D}_\mu \Lambda^A  - \theta_I^A \Lambda_\mu^I
\;,\nn\w2
\delta_{{\rm cov,}\,\Lambda} B_{\mu\nu}^{I} &=&
2\,{\cal D}_{[\mu}\Lambda_\nu^{I}
-2\Gamma^I_{AB}\Lambda^{(A} {\cal G}_{\mu\nu}^{B)}
-  \,\theta^{AI} \,\Lambda_{\mu\nu\,A}\ ,
\nn\w2
\delta_{{\rm cov,}\,\Lambda} C_{\mu\nu\rho\,A}
&=&
3\,{\cal D}_{[\mu}\Lambda_{\nu\rho]\,A}
 +6 \Gamma^{I}_{AB}\,{\cal G}_{[\mu\nu}^B\,\Lambda_{\rho]I}
+ 2 \Gamma^{I}_{AB}\,\Lambda^B\,{\cal H}_{\mu\nu\rho\,I}\ ,
\label{gauge}
\eea
with gauge parameters $\Lambda^A$, $\Lambda^I_{\mu}$, $\Lambda_{\mu\nu\,A}$, the covariant variation
$\delta_{cov.} B_{\mu\nu}^I$ as defined in \eqq{deltacov}, and
\bea
\delta_{\rm cov.} C_{\mu\nu\rho\,A}&\equiv&
\delta C_{\mu\nu\rho\,A}
-6\,\Gamma^{I}_{AB}\,B_{[\mu\nu\,I} \,\delta A_{\rho]}{}^B
- 2  \Gamma^{I}_{AB}(\Gamma_{I})_{CD}\,
A_{[\mu}^B A_{\nu}^C\,\delta A_{\rho]}^D\ .
\label{D1}
\eea
The modified Bianchi identities are given by
\bea
\cD_{[\mu}{\cal G} _{\nu\rho]}^A&=& \ft{1}{3}\theta_I^{A} {\cal H}^I_{\mu\nu\rho} \ ,
\label{B1}\w2
\cD_{[\mu} {\cal H}^I_{\nu\rho\sigma]} &=& \ft32 \Gamma^I_{AB}  \,{\cal G}^A_{[\mu\nu}  {\cal F}^B_{\rho\sigma]}
+\ft1{4} \theta^{IA}{\cal G}_{\mu\nu\rho\sigma\,A}\ .
\label{B2}
\eea
Just as consistency of the gauge algebra on the vector fields above gave rise to the constraint
(\ref{lincon1}), an analogous constraint follows from closure of the algebra on the two-forms:
\bea
(X_{A})_I{}^J &=& 2\left(\theta_I^B \Gamma^J_{AB} - \theta^{JB} (\Gamma_I)_{AB}\right)
\;.
\label{lincon2}
\eea
Otherwise a consistent gauge algebra would require the presence of
more than the (available) $n_V$ three-forms $C_{\mu\nu\rho\,A}$.
Recalling that the generator on the l.h.s.\ is defined by (\ref{generator}), this constraint
translates into the relation
\bea
\Theta_A{}^{IJ} &=& -\Gamma^{[I}_{AB}\,\theta^{J]B}_{\vphantom{I}}
\;,
\label{solTh}
\eea
between the various components of the embedding tensor.
Putting this together with (\ref{lincon1}) and the fact that $\Gamma^I_{AB}$
is an invariant tensor, one explicitly obtains the magical  $\Gamma$-matrix identity (\ref{magic}).
This shows in particular, that for values of $n_T$ different from $2, 3, 5, 9$, the non-abelian
gauge algebra does not close (in accordance with the appearance of the classical gauge anomaly in the action).\footnote{Note that this argument singling out once more the
magical cases does not even rely on the existence of a supersymmetric action.}
Furthermore, this calculation gives rise to the linear relation
\bea
\Gamma^I_{D C} X_{[AB]}{}^D  =
2 \Gamma^I_{D[A}\Gamma^J_{B]C}\theta_J^D
-\ft43  (\Gamma_J)^{\vphantom{I}}_{D[A} \Gamma^J_{B]C}\theta^{ID}
+\Gamma^I_{D[C} X^{\vphantom{I}}_{AB]}{}^D
\;.
\label{lincon3}
\eea
Using all the linear constraints
(\ref{lincon1}), (\ref{lincon2}), (\ref{lincon3}),
the quadratic constraint (\ref{quadconstraint}) finally translates into
the following set of relations
\bea
\theta^{IA}\eta_{IJ}\theta^{JB} = 0\;,
\quad
\theta^{IA} \Gamma^{[J}_{AB} \theta^{K]B} =0 \;,
\quad
X_{[AB]}{}^{C} \theta^{I B} &=& \Gamma^J_{AD}\, \theta_J^{C}\theta^{ID}
\;,
\nonumber\\[1.5ex]
{X_{[AB]}{}^D X_{[CD]}{}^E + X_{[CA]}{}^D X_{[BD]}{}^E
+ X_{[BC]}{}^D  X_{[AD]}{}^E}
 &=& \Gamma^I_{D[A}\, X^{\vphantom{I}}_{BC]}{}^D\, \theta_I^{E}
 \;.
 \label{quad2}
\eea
The first two relations turn out to be very restrictive.
In particular, the first equation implies that $\theta^{IA}$ is a matrix of mutually orthogonal null vectors
in $(n_T+1)$-dimensional Minkowski space, which requires that they are all
proportional, i.e.\ $\theta^{IA}$ factorizes as
\bea
\theta^{IA}=\zeta^A \xi^I\;,
\label{solth}
\eea
with an unconstrained (commuting) spinor $\zeta^A$ and $\xi^I \xi_I=0$\,.
The second equation of (\ref{quad2}) then has the unique solution
(up to irrelevant normalization)
$\xi^I =\Gamma^{I}_{AB}\zeta^A\zeta^B$
which defines a null vector by virtue of the identity (\ref{magic}).
The same identity implies that the tensor $\theta^{IA}$ is $\Gamma$-traceless:
\beq
\Gamma^I_{AB} \theta_I^B=0\ .
\label{tc}
\eeq
From these results one can already deduce some important facts on the structure of the gauge group.
Let us decompose the gauge group generators as
\beq
X_A=\hat{X}_A+\mathring{X}_A\ ,
\eeq
according to (\ref{generator}) into the part acting within the isometry group ${SO}(n_T,1)$ and the contribution of generators $t_{\cal X}$, respectively. Using (\ref{generator}), (\ref{solTh}) and (\ref{solth}) together with the $\Gamma$-matrix algebra and the magical identity (\ref{magic}) we find
\bea
\hat{X}_{(BC)}{}^A &=&
-\zeta^D\zeta^E\zeta^F\,(\Gamma_{IJ})_{(B}{}^A  \Gamma^{I}_{C)F} \Gamma^J_{DE}
\nonumber\\
&=&
\ft12\zeta^D\zeta^E\zeta^F\,(\Gamma_J)_{DE} (\Gamma^{I}\Gamma^J)_{F}{}^A (\Gamma_{I})_{BC}
~=~ \Gamma^I_{BC} \theta^{IA}
\;.
\eea
Comparing this to the linear constraint (\ref{lincon1}) thus implies $\mathring{X}_{(BC)}{}^A=0$. Some closer inspection then shows that this furthermore implies
\beq
\mathring{X}_{BC}{}^A=0\ .
\eeq
Thus, within the vector/tensor sector, the gauge group entirely lives within the scalar isometry group ${SO}(n_T,1)$, even in the cases $n_T=3$ and $n_T=5$, where the full global symmetry  group of the action possesses additional factors. All other equations of (\ref{quad2}) can then be shown to be identically satisfied.
Summarizing, the possible gaugings of the magical theories are entirely determined by the choice of a constant spinor $\zeta^A$ of the isometry group ${SO}(n_T,1)$. In the following we will discuss the possibility of inequivalent choices of $\zeta^A$ and subsequently study the structure of the resulting gauge group.

\subsection{Spinor orbits}

With the gauging determined by the choice of a spinor, one may wonder whether there are
different orbits of the action of $SO(n_T,1)$ on the spinorial representation which would represent inequivalent gaugings. The orbits of spinors up to 12  dimensions were studied long ago by Igusa \cite{MR0277558}.  More recently the spinors in critical dimensions were studied by Bryant \cite{MR1822355,Bryant}, whose study uses heavily the connection between spinors in critical   dimensions and the four division algebras $\mathbb{R}, \mathbb{C},\mathbb{H}$ and $\mathbb{O}$.\\

Understanding of the structure of  gauge groups we obtain  as well as the connection to the work of Bryant on orbits are best achieved by studying the embedding of the symmetries of the $6D$ magical theories in the corresponding  $5D$ supergravity theories obtained by dimensional reduction. Hence we shall first  review briefly the $5D$ magical supergravity theories.

Ungauged magical supergravity theories in five dimensions are Maxwell-Einstein supergravities that describe the coupling of pure $N=2$ supergravity to 5, 8, 14 and 26 vector multiplets , respectively. They  are uniquely defined  by simple Euclidean Jordan
algebras , $J_3^\mathbb{A}$ , of degree three generated by $ 3\times 3$ Hermitian matrices  over the four division algebras $\mathbb{A}$ = $\mathbb{R}$,
$\mathbb{C}$, $\mathbb{H}$(quaternions), $\mathbb{O}$(octonions).
The vector fields in these theories , including the graviphoton, are in one-to-one correspondence with the elements of the underlying simple Jordan algebras.
Their scalar manifolds are symmetric spaces of the form:
\eq
\mathcal{M}_5(J_3^{\mathbb{A}}) = \frac{Str_0(J_3^{\mathbb{A}})}{Aut(J_3^{\mathbb{A}})}
\en
where $Str_0(J_3^{\mathbb{A}}) $ and  $Aut(J_3^{\mathbb{A}})$ are the reduced structure and automorphism group of $J_3^{\mathbb{A}}$ , respectively, which we  list below~\cite{Gunaydin:1983rk,Gunaydin:1983bi}
\begin{equation}
\begin{split}
\mathcal{M}_5 (J_3^{\mathbb{R}}) &= \frac{SL(3,\mathbb{R})}{SO(3)}  \cr
\mathcal{M}_5 (J_3^{\mathbb{C}}) &= \frac{SL(3,\mathbb{C})}{SU(3)}  \cr
\mathcal{M}_5 (J_3^{\mathbb{H}}) &= \frac{SU^*(6)}{USp(6)}  \cr
\mathcal{M}_5 (J_3^{\mathbb{O}}) &= \frac{E_{6(-26)}}{F_4}
\end{split}
\end{equation}
They can be truncated to
theories belonging to the so-called generic Jordan family generated by reducible Jordan algebras  ( $\mathbb{R} \oplus J_2^\mathbb{A} $) where $J_2^{\mathbb{A}}$ are the Jordan algebras generated by $2\times 2$ Hermitian matrices  over $\mathbb{A}$. The isometry groups of the scalar manifolds of the  $5D$ theories resulting from the truncation are as follows:
\begin{equation}
\begin{split}
    Str_0[ \mathbb{R} \oplus J_2^\mathbb{R}]  & = {SO}(1,1) \times {Spin}\left(2,1\right) \subset {SL}\left(3, \mathbb{R}\right) \cr
   Str_0[\mathbb{R} \oplus J_2^\mathbb{C}] & = {SO}(1,1) \times {Spin}\left(3,1\right) \subset {SL}\left(3, \mathbb{C}\right) \cr
     Str_0[ \mathbb{R} \oplus J_2^\mathbb{H}]  &= {SO}(1,1) \times {Spin}\left(5,1\right) \subset {SU}^\ast\left(6\right) \cr
     Str_0[ \mathbb{R} \oplus J_2^\mathbb{O}]  & = {SO}(1,1) \times {Spin}\left(9,1\right) \subset {E}_{6(-26)}
\end{split}
\end{equation}
These truncated theories descend from $6D$ supergravity theories with $n_T=2,3,5$ and $n_T=9$ tensor multiplets and no vector multiplets.
A general element of the  Jordan algebras $J_3^\mathbb{A}$ of degree three can be decomposed with respect to its  Jordan subalgebra $J_2^\mathbb{A}$ as
\eq
X = \left(\begin{array}{cc}J_2^{\mathbb{A}} & \psi(\mathbb{A}) \\\psi^{\dagger}(\mathbb{A}) & \mathbb{R}\end{array}\right)
\en
where $\psi(\mathbb{A})$ is a two component spinor over $\mathbb{A}$
\beq
\psi(\mathbb{A}) =\left(\begin{array}{c}q_1 \\q_2\end{array}\right)
\eeq
and $\dagger$ represents transposition times conjugation in the underlying division algebra $\mathbb{A}$. Using this decomposition it was shown in \cite{Sierra:1986dx} that the Fierz identities for supersymmetric Yang-Mills theories in critical dimensions follow from the adjoint identities satisfied by the elements of $J_3^\mathbb{A}$ that define the magical supergravity theories in five dimensions \cite{Gunaydin:1983rk,Gunaydin:1983bi}. In the $6D$ magical supergravity theories the tensor fields correspond to the elements of $J_2^{\mathbb{A}}$ and the vector fields are represented by the elements\footnote{ The singlet vector field of the $5D$ theory represented by $\mathbb{R}$ corresponds to the vector field that comes from the $6D$ graviton. The bare graviphoton of the $5D$ Maxwell-Einstein supergravity is a linear combination of this vector field and the vector field that descends from the gravitensor of the $6D$ theory corresponding to the identity element of the Jordan algebra of degree three.}
\beq
\left(\begin{array}{cc}0 & \psi(\mathbb{A}) \\\psi^{\dagger}(\mathbb{A}) & 0\end{array}\right)
\eeq
The isometry group of the scalar manifold of a $6D$ magical supergravity is given by the reduced structure group $Str_0(J_2^{\mathbb{A}})$ of $J_2^{\mathbb{A}}$. They are well known to be isomorphic to the linear fractional groups
$SL(2,\mathbb{A}) $
for $\mathbb{A} = \mathbb{R}, \mathbb{C}, \mathbb{H}$ :
\begin{equation}
\begin{split}
Str_0(J_2^{\mathbb{R}}) &= Spin(2,1) = SL(2,\mathbb{R}) \\
Str_0(J_2^{\mathbb{C}}) &= Spin(3,1) = SL(2,\mathbb{C}) \\
Str_0(J_2^{\mathbb{H}}) &= Spin(5,1) = SL(2,\mathbb{H})
\end{split}
\end{equation}
The isometry group $Spin(9,1)$ of the octonionic theory can be similarly interpreted using the Jordan algebraic formulation
\eq
Str_0(J_2^{\mathbb{O}})= Spin(9,1) = SL(2,\mathbb{O})
\en

Orbits of the spinors appearing in Table 1 corresponding to the vector fields of $6D$ magical  supergravity theories under the action of the isometry groups  $Str_0( J_2^{\mathbb{A}})$ of their scalar manifolds were studied by Bryant using their realizations as  two component spinors $\psi ( \mathbb{A})$ over the underlying division algebras $\mathbb{A}$.
According to Bryant the entire spinor space $\psi(\mathbb{O})$  corresponding to a Majorana-Weyl  spinor forms a single orbit under the action of $Spin(9,1)$ with the isotropy  group $Spin(7)\circledS T_{8}$ \cite{MR1822355,Bryant}:
\eq
Orbit( \psi ({\mathbb{O}})) = \frac{Spin(9,1)}{Spin(7)\circledS T_8}
\label{orbit9}
 \en
where $T_8$ denotes the  eight dimensional translations.
Similarly, the quaternionic spinor $\psi ({\mathbb{H}})$ corresponding to a symplectic Majorana-Weyl spinor forms a single orbit under the action of $Spin(5,1)$:
\eq
Orbit( \psi ({\mathbb{H}})) = \frac{Spin(5,1)}{SU(2)\circledS T_4} \en
Complex two component spinor $\psi(\mathbb{C})$ is a Weyl spinor and forms a single orbit of $SL(2,\mathbb{C}) = Spin(3,1)$
 \eq
Orbit( \psi ({\mathbb{C}})) = \frac{SL(2,\mathbb{C})}{ T_2} \en
Restricting to real spinors one finds
\eq
Orbit( \psi ({\mathbb{R}})) = \frac{SL(2,\mathbb{R})}{ T_1} \en

To summarize, in all cases there is a single spinor orbit, such that different choices of the spinor $\zeta^A$ lead to equivalent gaugings.


\section{Structure of the gauge group and new couplings}

\subsection{The gauge group in the vector/tensor sector}

With the above results, the gauge group generators (\ref{generator}) in the vector/tensor sector take the explicit form
\bea
(\hat{X}_{A})_{B}{}^C &\!=\!&  (\bar\zeta\Gamma^{I}\zeta) (\Gamma^J\zeta)_A (\Gamma_{IJ})_B{}^C
\;,\nonumber\\[1.5ex]
(\hat{X}_A)^{IJ}  &\!=\!& 4\,
(\bar\zeta \Gamma^{[I} \zeta)(\Gamma^{J]}\zeta)_A
\;,
\label{generatorSO}
\eea
in terms of the spinor $\zeta^A$.
A simple calculation shows that
\bea
(\hat{X}_A \hat{X}_B)_C{}^D &=& 0
\;,\nonumber\\
(\hat{X}_A \hat{X}_B \hat{X}_C)_I{}^J &=& 0
\;,
\eea
i.e.\ these generators span an
$(n_T-1)$-dimensional nilpotent abelian algebra\footnote{
The cubic nilpotency of the generators in the vector representation
can also be seen by identifying this representation in the tensor product of two spinor representations.
}
and gauge  $(n_T-1)$ translations.
Remarkably,  this seems to be the only possible gauge group.
{\it Furthermore  these $(n_T-1)$ translations  can not lie strictly within the isometry group
 $\mathfrak{so}(n_T,1)$, cf.~(\ref{Ntranslations}) due to the appearance of central extensions of the gauge algebra as we explain below.}

Having seen above, that there is a single spinor orbit, all different choices of the spinor $\zeta^A$
lead to equivalent gaugings and it will be useful to give a presentation of the generators (\ref{generatorSO})
in an explicit basis.
We will proceed with the analysis of the maximal case $n_T=9$ from which all the lower magical theories can be obtained by truncation.
For $SO(9,1)$, according to (\ref{orbit9}) the compact part of the little group of a spinor is an $SO(7)$ under which the fundamental representations decompose as
\bea
\zeta^A: {\bf 16}_c &\rightarrow& {\bf 8_++7_-+1_-} \;,\qquad
\xi^I: {\bf 10} ~\rightarrow~ {\bf 8_0+1_{+2}+1_{-2}}
\;,\nonumber\\
A &\rightarrow& (\alpha, {t}, 0) \;,
\qquad\qquad\quad\,
I ~\rightarrow~ (\alpha, +,-) \;,
\label{defbasis}
\eea
with the subscripts referring to the $SO(1,1)$ charges in the decomposition (\ref{Ntranslations}).
This basis corresponds to the spinor $\zeta^A$ pointing in a given direction $\zeta^A = \frac12(\vec{0},\vec{0},g^{1/3})$,
such that the gauge group generators (\ref{generatorSO}) take the explicit form
\bea
\mbox{in the {\bf 16}}&:&
(\hat{X}_\alpha)_\beta{}^{t} = g\gamma^{t}_{\alpha\beta} \;,\quad
(\hat{X}_{\alpha})_\beta{}^0 = g\delta_{\alpha\beta}
\;,
\nonumber\\[1ex]
\mbox{in the {\bf 10}}&:&
(\hat{X}_{\alpha})_\beta{}^- = -(\hat{X}_{\alpha})_+{}^\beta =  g\delta_{\alpha\beta}
\;,
\label{Xexp}
\eea
with (antisymmetric) $SO(7)$ gamma matrices $\gamma^{t}_{\alpha\beta}$ ($\, t=1,\dots ,7$),
and all other components vanishing. We have introduced an explicit coupling constant $g$
that carries charge $-3$ under $SO(1,1)$\,.

In this basis, the full non-semisimple (nilpotent) structure of the gauge algebra (\ref{quadconstraint})
becomes explicitly:
\bea
[X_\alpha , X_\beta ] = -g \gamma^t_{\alpha\beta}\,X_t
\;,\qquad
[X_\alpha, X_t] =0= [X_t, X_u]
\;,\qquad
X_0=0
\;.
\label{algebra}
\eea
The generators $X_t$ thus act as central extensions of the algebra which
vanish when evaluated on vector or tensor fields: $(X_t)_I{}^J = 0 = (X_t)_A{}^B$, but which may
have a non-trivial action in the hypermultiplet sector, as we shall discuss in section~ 4.2\footnote{We should point out that a similar phenomenon of central extension of gauge groups arises also in $4D$ supergravity theories obtained by dimensional reduction
of  $5D$, $N=2$ Yang-Mills -Einstein supergravity theories coupled to tensor fields\cite{Gunaydin:2005bf}. }.

The structure of the centrally extended Abelian nilpotent gauge group is best understood by studying the embedding of the gauge group into the U-duality group of the corresponding ungauged $5D$ Maxwell-Einstein supergravity. For the exceptional supergravity $5D$ U-duality group is $E_{6(-26)}$ whose Lie algebra has a 3-graded decomposition with respect to the isometry group $SO(9,1)$ of $6D$ theory:
\eq
E_{6(-26)} = K_{\bf 16_{\bf c}} \oplus  SO(9,1)\times SO(1,1)_D \oplus T_{\bf 16_{\bf c}}
\en
where $T_{\bf 16_{\bf c}}$ denotes the 16 dimensional translational symmetries corresponding to Abelian gauge symmetries of the vector fields of $6D$ theory.
 The generator $\Delta$ that determines the 3-grading of $SO(9,1)$ ( see (\ref{Ntranslations}) ) leads to a 5-grading of $E_{6(-26)}$ so that $D$ and $\Delta$ determine a 5 by 3 grading of $E_{6(-26)}$ with respect to its $SO(8)\times SO(1,1)_{\Delta}\times SO(1,1)_{D} $ subgroup as shown in Table \ref{5by3grading}.

\begin{table}
\begin{equation} \nonumber
\begin{array}{ccccc}
 \phantom{R} & T_{\bf{8^c}} & \vline\phantom{K} & T_{\bf{8^s}} & \phantom{R}  \\[20pt]
 \tilde{N}_{\bf{8^v}} &---- \phantom{U} & \left( SO(8) \times SO(1,1)_{\Delta} \times SO(1,1)_{D}  \right) &----             \phantom{V} & N_{\bf{8^v}}  \\[20pt]
 \phantom{R} & K_{\bf{8^s}} & \vline\phantom{K} & K_{\bf{8^c}} & \phantom{R}\\[8pt]
\end{array}
\end{equation}
\caption{ \label{5by3grading} Above we give the 5 by 3 grading of $E_{6(-26)}$ with respect to the generators $ \Delta$ and $D$ respectively. Eight dimensional representations that are in triality are denoted as  ${\bf 8^v}, {\bf 8^c}$ and ${\bf 8^s}$.}
\end{table}
Restricting to the $Spin(7)$ subgroup such that $SO(8)$ irreps decompose as
\eqn
{\bf 8^v = 8} \\ \nn
{\bf 8^s =7+1 } \\
{\bf 8^c = 8 } \nn
\enn
one finds that the 8 generators
$ (T_\alpha + N_\alpha )$ , transforming in the spinor representation of $Spin(7)$, form a centrally extended  nilpotent Abelian subalgebra with 7 generators $ T_t $ acting as its central elements.  They generate a nilpotent subgroup of  $F_{4(-20)}$ which  is a subgroup of  $5D$ U-duality group $E_{6(-26)}$. $F_{4(-20)}$  admits a 5-grading of the form:
\eq
F_{4(-20)} ={\bf  7_{-2} \oplus  8_{-1} } \oplus ~ Spin(7) \times SO(1,1) {\bf  \oplus ~ 8_{+1} \oplus 7_{+2}}
\;.
\label{dec8}
\en
Thus the generators that gauge the centrally extended Abelian subalgebra in the embedding tensor formalism can be uniquely identified with the generators of this
nilpotent subalgebra of $F_{4(-20)}$
\begin{equation}
X_\alpha \equiv  T_\alpha + N_\alpha\ ,\qquad   X_t  \equiv  T_t\ ,
\end{equation}
where $\alpha = 1,2,\cdots, 8$ and $ t=1,2,\cdots ,7 $.

The quaternionic $6D$ magical supergravity has $SU^*(6)$ as its $5D$ U-duality group which has a 3-grading with respect to its $SU^*(4)\times SU(2)$ subgroup.
The vector fields of the $6D$ theory transform as a symplectic Majorana-Weyl spinor of $SU^*(4)\times SU(2)$.  The analog of $F_{4(-20)}$ is the $USp(4,2)$ subgroup of $SU^*(6)$. The centrally extended nilpotent Abelian translation gauge group sits inside  $USp(4,2)$  which has the 5-graded decomposition
\eq
USp(4,2)= {\bf 3_{-2} \oplus 4_{-1}} \oplus SU(2)\times SO(1,1) \oplus {\bf 4_{+1} \oplus 3_{+2}}
\;.
\label{dec4}
\en

As for the complex magical theory the $5D$ U-duality group $SL(3,\mathbb{C}) $ has a 3-grading with respect to its $SL(2,\mathbb{C})\times U(1)$ under which the vector fields of $6D$ theory transform as a pair of complex Weyl spinors. The nilpotent gauge group sits inside the $SU(2,1)$ subgroup of $SL(3,\mathbb{C})$ and has the 5-grading
\eq
SU(2,1) = {\bf 1_{-2}\oplus 2_{-1}} \oplus U(1)\times SO(1,1) \oplus  {\bf 2_{+1} \oplus 1_{+2}}
\;.
\label{dec2}
\en

The simple groups in which the centrally extended Abelian gauge groups of octonionic, quaternionic and complex magical theories can be minimally embedded
satisfy the following chain of inclusions
\eq
F_{4(-20)} \supset USp(4,2) \times USp(2) \supset SU(2,1) \times U(1)
\en

The fact that the method of embedding tensor formalism leads to unique centrally  extended  Abelian gauge groups for each of the magical supergravity theories is quite remarkable. Even though the ``central charges" act trivially  on the vector and tensor fields as must be evident from the above analysis they  may have nontrivial action on the hyperscalars  as will be discussed in the next section \footnote{  We should note that the gaugings we presented are quite different from Scherk-Schwarz type gaugings\cite{Andrianopoli:2002mf}, obtained from a  3-grading of the  symmetry group of a lower dimensional theory,  which always include a generator from the grade zero subalgebra whereas ours do not. The gaugings above involve a subtle interplay between 3-grading and 5-grading and allow for central charges. Furthermore Scherk-Schwarz gaugings  are obtained from a higher dimensional theory and our gaugings do not have higher dimensional origins.}.

\subsection{The gauge group in the hypersector}
\label{sec:gaugehypers}

We will now also allow for isometries of the quaternionic K\"ahler manifold to be gauged,
i.e.\ consider the full generator (\ref{generator}) with non-vanishing $\Theta_A{}^{\cal A}$. The generators $t_{{\cal A}}$ denote the
isometries on the quaternionic K\"ahler manifold acting by a Killing vector field $K^X_{{\cal A}}$
\bea
t_{\cal A}\cdot \phi^X &=& K^X_{{\cal A}}(\phi)
\;,
\eea
on the hyperscalars. It is straightforward to derive that under this transformation the
$Sp(1)_R \times Sp(n_H)$ connections transform as
\bea
t_{\cal A}\cdot Q_\mu{}_i{}^j &=& \partial_\mu \phi^X D_X ({\cal S}_{\cal A}){}_i{}^j\;,\qquad
t_{\cal A}\cdot Q_\mu{}_r{}^s ~=~ \partial_\mu \phi^X D_X ({\cal S}_{\cal A})_r{}^s\;,
\eea
with
\bea
({S}_{\cal A})_i{}^j &\equiv & K_{\cal A}^X A_X{}_i{}^j + C_{\cal A}{}_i{}^j
\;,\qquad
({S}_{\cal A})_r{}^s ~\equiv ~ K_{\cal A}^X A_X{}_r{}^s + C_{\cal A}{}_r{}^s
\;,
\label{defS}
\eea
in terms of the functions $C_{\cal A}{}_i{}^j$ and $C_{\cal A}{}_r{}^s$ defined in (\ref{defC}), (\ref{Crs}), respectively, for the Killing vector field $K_{\cal A}$\,.
From this, we conclude that the fermion fields transform as
\bea
t_{\cal A}\cdot \chi^a_i &=& -({S}_{\cal A})_i{}^j\,\chi^a_j
\;,\qquad
t_{\cal A}\cdot \psi_r ~=~ -({S}_{\cal A})_r{}^s\,\psi_s\;,
\qquad
\mbox{etc.}
\label{trafS}
\eea
Upon gauging, the gauge covariant derivative of the hyperscalars is given by
\beq
{\cal D}_\mu \phi^X = \partial_\mu \phi^X -g A_\mu^A {\cal K}_A^X\ ,\qquad {\cal K}_A^X \equiv \Theta_A{}^{\cal A} K_{\cal A}^X\ ,
\eeq
and the gauge covariant derivatives of the fermion fields by
\bea
{\cal D}_\mu \psi_\nu^i &=&
\nabla_\mu \psi_\nu^i + {\cal Q}_\mu{}^{ij} \psi_{\nu j}\ ,
\nn\w2
{\cal D}_\mu \chi^{ai} &=&
\nabla_\mu \chi^{ai} + {\cal Q}_\mu^{ab} \chi_b^i + {\cal Q}_\mu{}^{ij} \chi_j^a\ ,
\nn\w2
{\cal D}_\mu \psi^r &=& \nabla_\mu \psi^r + {\cal Q}_\mu{}^{rs} \psi_s\ ,
\nn\w2
{\cal D}_\mu \lambda^{Ai} &=& \nabla_\mu \lambda^{Ai} + {\cal Q}_\mu{}^{ij} \lambda^A_j
-A_\mu^C X_{CB}{}^A\,\lambda^{Bi}\ ,
\label{connections}
\eea
with\footnote{See \cite{Percacci:1998ag} for a description of the ${\cal S}$-functions in the context of
$G/H$ coset sigma models in which an arbitrary subgroup of $G$ is gauged. }
\bea
\cQ_\mu^{ab} &=& L^{I[a}\,\cD_\mu L_I^{b]}\ ,
\label{bb1}\w2
{\cal Q}_\mu{}^{ij} &=& {\cal D}_\mu \phi^X A_X{}^{ij}
+ A_\mu^A {\cal S}_A{}^{ij}
\nn\w2
&=&  \partial_\mu \phi^X A_X{}^{ij}
+ A_\mu^A {\cal C}_A{}^{ij}  \ ,
\label{bb2}\w2
{\cal Q}_\mu{}^{rs} &=& {\cal D}_\mu \phi^X A_X{}^{rs}
+ A_\mu^A {\cal S}_A{}^{rs}
\nn\w2
&=&  \partial_\mu \phi^X A_X{}^{rs}
+ A_\mu^A {\cal C}_A{}^{rs} \ ,
\label{bb3}
\eea
with the following definitions
\beq
\begin{split}
{\cal S}_A{}^{ij} &= {\cal K}_A^X A_X{}^{ij} + {\cal C}_A{}^{ij}\ ,
\w2
{\cal C}_{A i}{}^{j} &= -\frac{1}{2n_H}\, V^X_{ri} V_Y^{rj} \nabla_X {\cal K}_A^Y\ ,
\label{defCS}
\end{split}
\eeq
and similarly
\beq
\begin{split}
{\cal S}_A{}^{rs} &= {\cal K}_A^X A_X{}^{rs} + {\cal C}_A{}^{rs}\ ,
\w2
{\cal C}_A{}^{rs} &= -\ft12 V^X_{ri} V_Y^{si} \nabla_X {\cal K}_A^Y \ .
\end{split}
\eeq
The constraint analysis in the vector/tensor sector remains unchanged in presence of the $t_{{\cal A}}$, such that the gauge group in this sector still reduces to the set of $(n_T-1)$ Abelian translations as we have derived above. Gauge invariance of
the new components $\Theta_A{}^{\cal A}$ of the embedding tensor on the other hand implies that
\bea
[ \Theta_A{}^{\cal A} t_{{\cal A}} , \Theta_B{}^{\cal B} t_{{\cal B}} ] &=& -X_{AB}{}^C\,\Theta_C{}^{\cal A} t_{{\cal A}}
\;,
\label{algebrahypers}
\eea
with the same structure constants $X_{AB}{}^C$ encountered in (\ref{quadconstraint}). It follows from
(\ref{quadconstraint}) that
\beq
\theta^{IA}\, \Theta_A{}^{\cal A}=0\ .
\label{tt}
\eeq
Furthermore, using the explicit form (\ref{algebra}) of the structure constants $X_{AB}{}^C$, we find that $\Theta_0{}^{\cal A}=0$ and the gauging in the hypersector of the octonionic magical supergravity corresponds to selecting 8+7 Killing vector fields $\KK_\alpha\equiv\Theta_\alpha{}^{\cal A} K_{\cal A}$, $\KK_t\equiv\Theta_t{}^{\cal A} K_{\cal A}$ (not necessarily linearly independent), which satisfy the algebra
\bea
[\KK_\alpha , \KK_\beta ] = -g \gamma^t_{\alpha\beta}\,\KK_t
\;,\qquad
[\KK_\alpha, \KK_t] =0= [\KK_t, \KK_u]
\;.
\label{algebraKV}
\eea
Thus, the generators associated with the full gauge group in the magical supergravities, including both the vector-tensor and hyper sectors, are
\beq
X_A = \{ {\widehat X}_\alpha +{\cal K}_\alpha\ ,{\cal K}_t \}\ ,
\eeq
The existence of a combination of Killing vectors that satisfy this algebra in its maximal form, i.e. with none of the generators set to zero, is a nontrivial constraint, since such an algebra does not necessarily lie in the isometry of the hyperscalar manifold.  A trivial solution to the constraints \eqq{algebraKV} is given by setting $\KK_\alpha=0=\KK_t$ in which case the gauge group simply does not act in the hypersector, and has the generators $X_\alpha$. A less trivial option is the choice $\KK_t=0$ in which case the gauge algebra generators consist of ${\widehat X}_\alpha +{\cal K}_\alpha$, and $\KK_\alpha$ can be chosen to be any set of up to 8 commuting (compact, noncompact or nilpotent) isometries, which may in particular include $U(1)_R$ subgroup of the $Sp(1)_R$ R-symmetry group.

In general, and in contrast to the vector/tensor sector, in the hypersector the generators $\KK_t$ may act as nontrivial central charges of the gauge algebra. We can solve the constraints
(\ref{algebraKV}) by selecting an ideal ${\cal I}$ inside the algebra
${\cal A}$ defined by (\ref{algebraKV}), representing all generators in ${\cal I}$ by zero, and
embedding the quotient ${\cal A}/{\cal I}$ into the isometry algebra of the quaternionic manifold.
(The solution considered above where ${\cal K}_t$ is set to zero is a particular example of this procedure). In this case the generators will be embedded among the positive root generators.
For a coset manifold and its representative in the corresponding triangular gauge,
their action does not induce a compensating transformation acting on the fermions,
implying that the matrices $({\cal S}_A)_i{}^j$, $({\cal S}_A)_r{}^s$ of (\ref{defS}), (\ref{trafS}) vanish.

As an illustration we give some examples of embeddings of the nilpotent gauge groups with nontrivial central charges into simple quaternionic Lie groups
\begin{itemize}

\item
As we discussed  above, the nilpotent gauge algebra  (\ref{algebraKV}) of the octonionic magical theory  with all seven $\KK_t$ non-vanishing can be embedded into
the Lie algebra of the group $F_{4(-20)}$ which admits a five grading according to (\ref{dec8})
\bea
{\bf {7}_{-2} \oplus {8}_{-1} } \oplus Spin(7)_0\oplus SO(1,1)_0 \oplus  {\bf {8} _{+1} \oplus {7}_{+2}}
\;,
\label{emb2}
\eea
with the obvious embedding of (\ref{algebraKV}) as the generators of positive grading.
The group $F_{4(-20)}$ may be embedded into the isometry group of the
coset space $E_{6(-26)}/F_4$ and via the chain along the first line of table~\ref{tab:cosets} further into the
isometry group of the quaternionic K\"ahler manifold $E_{8(-24)}/(E_7\times SU(2))$\,.
With hyperscalars in this particular quaternionic K\"ahler manifold, the algebra (\ref{algebraKV}) can thus be  realized.
Interestingly, this manifold is precisely the moduli space of this magical theory without hypers upon dimensional reduction
to $D=3$ dimensions, cf.~table~\ref{tab:cosets}. The corresponding $6D$ theory is anomaly free as we will discuss later and  the scalar manifold of the resulting ungauged $3D$ theory is doubly exceptional
\eq
\mathcal{M}_3 =  [\frac{E_{8(-24)}}{(E_7\times SU(2))}] \times [\frac{E_{8(-24)}}{(E_7\times SU(2))}]
\en

The quaternionic symmetric space of minimal dimension whose isometry group has $F_{4(-20)}$ as a subgroup is
\eq
\frac{E_{7(-5)}}{SO(12)\times SU(2)}
\en
However, $E_{7(-5)}$ does not have $E_{6(-26)}$ as a subgroup and the $6D$ octonionic magical theory coupled to hypermultiplets with this target manifold  is not anomaly free.

\item
An example of a non-maximal realization of (\ref{algebraKV}) for the octonionic magical theory,
is given by selecting one of the central charges, i.e.~splitting $\{\KK_t\} = \{\KK, \KK_{\tilde{t}}\}$
and setting the ideal ${\cal I}$ spanned by the six $\KK_{\tilde{t}}$ to zero. According to the structure of the $SO(7)$ gamma matrix in the structure constants (upon breaking $SO(7)$ down to the $SO(6)$
defined by $\KK$) the quotient ${\cal A}/{\cal I}$ is given by the algebra
\bea
[\KK_a, \KK_b] &=& 0 ~=~ [\KK^a, \KK^b]\;,\qquad
[\KK_a, \KK^b] ~=~ g \delta_a^b\,\KK
\;.
\label{sub1}
\eea
This algebra can e.g.\ easily be embedded into the quaternionic K\"ahler manifold
$SU(4,2)/S(U(4)\times U(2))$, whose isometry group admits a five grading according to
\bea
{\bf {1}_{-2} \oplus ( {4} + \bar{4})_{-1}} \oplus
U(3,1)_0\oplus O(1,1)_0 \oplus {\bf  ( {4} + \bar{4})_{+1} \oplus {1}_{+2}}
\;,
\label{emb1}
\eea
with the obvious embedding of (\ref{sub1}) as the generators of positive grading.

\item
According to (\ref{dec4})
the nilpotent gauge algebra of the quaternionic magical theory can be embedded into the Lie algebra of $USp(4,2)$ which is quaternionic real. Therefore the quaternionic symmetric space  of minimal dimension  whose isometry group includes the nilpotent gauge group with all three central charges is
\eq
\frac{USp(4,2)}{USp(4)\times USp(2)}
\en
If we require the isometry group of the hypermanifold to have the corresponding $5D$ isometry group $SU^*(6)$ as a subgroup we can follow the chain along the second line of table~\ref{tab:cosets} further to the target manifold
\beq
 \frac{E_{7(-5)}}{SO(12)\times SU(2)}
\eeq
The scalar manifold of the ungauged quaternionic magical theory coupled to this hypermatter has the double exceptional isometry group
\beq
 \frac{E_{7(-5)}}{SO(12)\times SU(2)}\times
 \frac{E_{7(-5)}}{SO(12)\times SU(2)}
\eeq
in three dimensions.

\item For the complex magical theory according to (\ref{dec2}) the nilpotent gauge group embeds into $SU(2,1)$ which is quaternionic real. Thus the minimal hypermanifold in this case is
\beq
\frac{SU(2,1)}{U(2)}
\eeq
Going along the third row of  table~\ref{tab:cosets} we can also couple the theory to hypermultiplets  with the target manifold
\beq
\frac{E_{6(2)}}{SU(6)\times SU(2)}
\eeq
Again the corresponding $3D$ target space with this hypersector is doubly exceptional.

\end{itemize}

\subsection{Gauging of R-symmetry}


In presence of hypermultiplets, the R-symmetry $Sp(1)_R$ is embedded into the isometries on
the quaternionic K\"ahler manifold and its gauging is a particular case of the construction discussed
in the previous section. In absence of hypermultiplets, the $R$-symmetry acts exclusively on the
fermions and may be included in the gauging by extending the gauge group generators (\ref{generator}) to
\bea
{X}_A &=&  \hat{X}_A + \Theta_A{}^{ij}  t_{ij} ~\equiv~  \hat{X}_A + \xi_A
\;,
\label{generatorR}
\eea
with $t_{ij}$ representing the $Sp(1)_R$ generators. In complete analogy to the
calculation leading to (\ref{algebrahypers}) one derives the conditions
\bea
[ \xi_A{} , \xi_B ] &=& -X_{AB}{}^C\,\xi_C
\;,
\eea
from which we find that the most general gauging in absence of hypermultiplets is given by
\bea
{X}_\alpha &=& \hat{X}_\alpha + \xi_\alpha c^{ij} t_{ij}
\;,
\qquad
{X}_t ~=~ 0 ~=~ {X}_0\;,
\label{abelian}
\eea
with $\hat{X}_\alpha$ from (\ref{Xexp}) and constant $\xi_\alpha$ and
$c^{ij}$ selecting a $U(1)$ generator within $Sp(1)_R$\,.
All formulas of the previous section apply, in particular the
connections on the fermion fields are still given by (\ref{connections}),
upon setting all hyperscalars to zero and with constant $C_{\alpha i}{}^j \equiv \xi_\alpha c_i{}^{j}$\,.
The gauge algebra is still of the form (\ref{algebra}) but with $X_t$ set to zero, i.e. $[X_\alpha,X_\beta]=0$, with $X_\alpha$ from  \eqq{abelian}.


\subsection{Gauged Magical Supergravities}


Putting together the ingredients described in previous sections, and following the standard Noether procedure,
we find that the action for gauged magical supergravity, up to quartic fermion terms, is given by
\beq
\begin{split}
e^{-1} {\cal L} &= R -\frac{1}{12}  g_{IJ} {\cal H}_{\mu\nu\rho}^I {\cal H}^{\mu\nu\rho J} -\frac14 {\cal P}_\mu^a {\cal P}^{\mu a}
-\frac12 {\cal P}_\mu^{ri} {\cal P}_{\mu ri}
-\frac14 m_{AB} {\cal G}_{\mu\nu}^A {\cal G}^{\mu\nu B}
\w2
& +\frac12 {\bar\psi}_\mu\gamma^{\mu\nu\rho} {\cal D}_\nu\psi_\rho
 -\frac12 {\bar\chi}^a\gamma^\mu {\cal D}_\mu \chi^a-\frac12 \bar{\psi}^r \gamma^\mu {\cal D}_\mu \psi_r
\w2
&
- m_{AB} {\bar\lambda}^A \gamma^\mu {\cal D}_\mu \lambda^B +\frac12{\bar\psi}_\mu\gamma_\nu\gamma^\mu\chi^a {\cal P}^\nu_a
-(\bar{\psi}_\mu^i \gamma_\nu\gamma^\mu \psi^r) {\cal P}^\nu_{ri}
\w2
& +\frac{1}{48} {\cal H}_{\mu\nu\rho} \left( {\bar\psi}^\lambda\gamma_{[\lambda}\gamma^{\mu\nu\rho}\gamma_{\tau]}\psi^\tau
+{\bar\chi}^a\gamma^{\mu\nu\rho}\chi^a-{\bar\psi}^r\gamma^{\mu\nu\rho}\psi_r
\right)
\w2
& +\frac{1}{24} {\cal H}_{\mu\nu\rho}^a\left(
\bar{\psi}_\lambda\gamma^{\mu\nu\rho}\gamma^\lambda\chi^a
- m_{a AB} \bar\lambda^A \gamma^{\mu\nu\rho} \lambda^B\right)
\w2
&-\frac12 {\cal G}_{\mu\nu}^A \left( m_{AB}\,{\bar\psi}_\lambda \gamma^{\mu\nu}\gamma^\lambda \lambda^B - m_{a AB}\, {\bar\chi}^a \gamma^{\mu\nu} \lambda^B\right)
+{\cal L}_{\rm top}  +  {\cal L}_{\rm Yukawa}  + {\cal L}_{\rm pot}\ ,
\label{action-1}
\end{split}
\eeq
where ${\cal L}_{\rm top}$ is the gauge invariant completion of the $B\wedge F\wedge F$ term and
its variation is
\bea
\delta {\cal L}_{\rm top} &=&
\ft16 \epsilon^{\mu\nu\rho\sigma\lambda\tau}\,\Gamma^I_{AB}
\left(
{\cal H}_{\mu\nu\rho\,I} {\cal G}^A_{\sigma\lambda}\,\delta_{\rm cov} A^B_{\tau}
-\ft34 {\cal G}_{\mu\nu}^A {\cal G}_{\rho\sigma}^B \, \delta_{\rm cov} B_{\lambda\tau\,I}
\right)
\nonumber\\
&&{}
+\ft1{48} \epsilon^{\mu\nu\rho\sigma\lambda\tau}\,
\theta^A_I \left(
{\cal G}_{\mu\nu\rho\sigma\,A} \,\delta_{\rm cov} B_{\lambda\tau\,I}
-\ft{4}{3}
{\cal H}^I_{\mu\nu\rho}\, \delta_{\rm cov} C_{\sigma\lambda\tau\,A} \right)
\;.
\eea
and the Yukawa couplings and the scalar potential are given by\footnote{For the $n_T=5$ theory
and in absence of hypermultiplets these couplings
have been obtained in~\cite{Roest:2009sn} by truncation from gaugings of the
maximal theory~\cite{Bergshoeff:2007ef}.}

\bea
e^{-1}{\cal L}_{\rm Yukawa} &=&
 \bar{\psi}^i_\mu \gamma^\mu \lambda^A_i\,  \theta^{IB} m_{AB} L_{I}
- \bar{\psi}_{\mu i} \gamma^\mu \lambda^A_j\,  \CC_{A}^{ij}
+\bar{\chi}^{ai} \lambda^A_i \, \theta^{IB}  L_{I} m_{a\,AB}
\nonumber\\[.8ex]
&&{}
- \bar{\chi}^a_{i} \lambda^A_j\,  m_{a\,AC} m^{CB}\,\CC_{B}^{ij}
-2 \bar{\psi}_r \lambda^A_i  V_X^{ri} \KK^X_A\ ,
\w2
e^{-1}{\cal L}_{\rm potential} &=&
- \frac14 \left( \theta^{IA} \theta^{JB} m_{AB} g_{IJ}
+  \CC_A{}_{ij} \CC_B^{ij} m^{AB} \right)
\;.
\eea
The functions  $g_{IJ}$, $m_{AB}$ and ${\cal C}_A^{ij}$ are defined in \eqq{defg}, \eqq{defMM} and \eqq{defCS}, respectively, and the gauge covariant derivatives of the fermions in \eqq{connections}.

The local supersymmetry transformations of the gauged theory, up to cubic fermions terms, are
\beq
\begin{split}
\delta e_\mu^m  &= \bar\eps\c^m\psi_\mu\ ,
\w2
\delta \psi_\mu &= {\cal D}_\mu \eps + \frac{1}{48}
\c^{\rho\sigma\tau} \c_\mu  \eps {\cal H}_{\rho\sigma\tau}\ ,
\w2
\delta_{\rm cov.} B_{\mu\nu}{}^I &= - 2 \bar\eps \c_{[\mu} \psi_{\nu]}\, L^I + \bar\eps\c_{\mu\nu} \chi^a\, L^I_a\ ,
\w2
\delta_{\rm cov.} C_{\mu\nu\rho\, A} &=  -\bar\eps \c_{\mu\nu\rho} \lambda_A\ ,
\w2
\delta \chi^a &= \frac12 \c^\mu \eps {\cal P}_\mu^a - \frac{1}{24} \c^{\mu\nu\rho} \eps
{\cal G}^a_{\mu\nu\rho}\ ,
\w2
\delta L_I &= \bar\eps\chi^a L_I^a\ ,
\w2
\delta A_\mu^A  &= \bar\eps \c_\mu \lambda^A \ ,
\w2
\delta\lambda^A &= - \frac14 \c^{\mu\nu} \eps {\cal G}^A_{\mu\nu}  -\ft12 \theta^{IA} L_{I}\, \epsilon_i
-\ft12  m^{AB} \CC_{B}{}_{ij}\,\epsilon^j \ ,
\w2
\delta\phi^X &= V^X_{ri}\,\bar{\epsilon}^{i} \psi^r\ ,
\w2
\delta\psi^r &= {\cal P}_\mu^{ri} \gamma^\mu\epsilon_i\ .
\label{t7-1}
\end{split}
\eeq
In establishing the supersymmetry of the action, it is important to recall that the
(anti)self-duality equations ${\cal H}_{\mu\nu\rho}^+=0$ and ${\cal H}_{\mu\nu\rho}^{a-}=0$ are to be
used after varying the action. In carrying out the Noether procedure, it is also useful to note that the gauge covariant scalar currents
\beq
\cP_\mu^a = L^{Ia} \cD_\mu L_I\ ,\qquad {\cal P}_\mu^{ri} = {\cal D}_\mu \phi^X V_X^{ri}\ ,
\label{msc}
\eeq
satisfy the relations
\bea
\cD_{[\mu}\cP_{\nu]}^a &=& -\frac12\, \cF_{\mu\nu}^A {\cal C}_A{}^a\ ,
\qquad {\cal C}_A{}^a = {\widehat X}_A^{IJ} L_I^a L_J\ ,
\label{dp1}\w2
{\cal D}_{[\mu} {\cal P}_{\nu]}^{ri} &=& -{\cal F}_{\mu\nu}^A {\cal C}_A^{ri}\ ,\qquad
 {\cal C}_A^{ri} = \Theta_A{}^{\cal A} K_{\cal A}^X V_X^{ri}\ .
 \label{dp2}
\eea
Also encountered in the Noether procedure are the curvatures associated with the connections defined in \eqq{bb1} and \eqq{bb2} which take the form
\bea
\cQ_{\mu\nu}^{ab} &=& -2\cP^a_{[\mu}\cP^b_{\nu]} -\cF_{\mu\nu}^A {\cal C}_A{}^{ab}\ ,
\qquad {\cal C}_A{}^{ab} = {\widehat X}_A^{IJ} L_I^a L_J^b\ ,
\label{curv1}\w2
{\cal Q}_{\mu\nu i}{}^j &=& 2{\cal P}_{[\mu}^{rj}{\cal P}_{\nu]ri} +  {\cal F}_{\mu\nu}^A {\cal S}_{Ai}{}^j\ ,
\label{curv2}
\eea
with ${\cal S}_{Ai}{}^j$ from \eqq{defCS}. In the course of Noether procedure, it is also useful to note that the curvature ${\cal F}_{\mu\nu}^A$ occurring in equations \eqq{dp1}--\eqq{curv2} can be replaced by ${\cal G}_{\mu\nu}^A$, by exploiting the constraint \eqq{tt}.

As emphasized above, the (anti)self-duality equations on the projections of ${\cal H}_{\mu\nu\rho}^I$ are to be imposed {\it after} the Euler-Lagrange variation of the action with respect to all fields. Indeed, the field equation for $C_{\mu\nu\rho A}$, which  has no kinetic term and it appears in the action only through ${\cal L}_{\rm top}$ and the ${\cal G}_{\mu\nu}^A$ dependent terms, is identically satisfied provided that the stated (anti)self-duality equations are used. Varying the action with respect to $B_{\mu\nu}^I$, on the other hand, again modulo the (anti)self-duality conditions and their consequences, give precisely the duality equation \eqq{de}. Another salient feature of the action is that the scalar potential is a positive definite expression. Using the explicit parametrization of $SO(n_T,1)$ given in (\ref{Lpar}) and the basis (\ref{Xexp}), the scalar potential takes the form
\bea
e^{-1}{\cal L}_{\rm potential} &=&
-\ft1{16} g^2  e^{-3\sigma}
- \ft18 e^{-\sigma}\left(\CC_\alpha^{ij} + \gamma^t_{\alpha\beta} \varphi^\beta \CC_t^{ij}\right)^2
-  \ft18 e^\sigma\, \CC_t{\,}_{ij} \CC_t^{ij}
\;,
\eea
with the functions $\CC_\alpha^{ij}$, $\CC_t^{ij}$ from (\ref{defC}) defined for the Killing
vector fields satisfying the algebra (\ref{algebraKV}). It follows immediately from the
$e^\sigma$ powers, that for $\CC_t^{ij}=0$ this potential does not admit extremal points.

We can gain more insight to the nature of the new couplings, by observing that in the normalization of generators given in (\ref{algebra}) the covariant field strengths (\ref{defG}) take the explicit form
\bea
{\cal G}^\alpha_{\mu\nu} &=&  2 \partial_{[\mu} A_{\nu]}^\alpha \;,\nonumber\\
{\cal G}_{\mu\nu}^{t} &=&
2\partial_{[\mu} A_{\nu]}^{t}
+ g \gamma^{t}_{\alpha\beta}  \,A_\mu{}^\alpha  A_\nu{}^\beta\;,\nonumber\\
{\cal G}_{\mu\nu}^0 &=& 2 \partial_{[\mu} A_{\nu]}^0 + g B^{+}_{\mu\nu}\ ,
\eea
and the minimal couplings of vectors to tensors are explicitly given by
\bea
{\cal D}_\mu B_{\nu\rho}^\alpha &=& \partial_\mu B_{\nu\rho}^\alpha +gA^\alpha_\mu B^+_{\nu\rho}
\ ,\nn\w2
{\cal D}_\mu B_{\nu\rho}^- &=& \partial_\mu B_{\nu\rho}^- -g A^\alpha_\mu B^\alpha_{\nu\rho}
\ ,
\nn\w2
{\cal D}_\mu B_{\nu\rho}^+ &=& \partial_\mu B_{\nu\rho}^+
\ .
\eea
In the explicit parametrization (\ref{Lpar}) and in the basis defined by (\ref{Xexp}),
the various components of the
kinetic matrices $g_{IJ}$ and $m_{AB}$ take the explicit form
\bea
g_{\alpha\beta} &=& \delta_{\alpha\beta}+2 e^{-2\sigma} \varphi_\alpha\varphi_\beta\;,\qquad
g_{\alpha-} ~=~ e^{-2\sigma} \varphi_\alpha\;,\qquad
g_{\alpha_+} ~=~ \left(1+ e^{-2\sigma} \varphi^\beta \varphi_\beta\right) \varphi_\alpha\;,
\nonumber\\
g_{+-} &=& e^{-2\sigma} \varphi^\alpha \varphi_\alpha\;,\qquad
g_{++} ~=~ \left(e^{\sigma}+e^{-\sigma} \varphi^\alpha \varphi_\alpha\right)^2\;,\qquad
g_{--} ~=~ e^{-2\sigma}
\;,\nonumber\\[1.5ex]
m_{\alpha\beta} &=&\ft12 \delta_{\alpha\beta} \left(e^{\sigma}+e^{-\sigma} \varphi^\alpha \varphi_\alpha\right)\;,\qquad
m_{tu} ~=~ \ft12e^{-\sigma} \delta_{tu}\;,\qquad
m_{00} ~=~ \ft12e^{-\sigma}\;,\nonumber\\
m_{t\alpha} &=& \ft12\gamma^t_{\alpha\beta}\,e^{-\sigma}  \varphi^\beta\;,\qquad
m_{0\alpha} ~=~ \ft12e^{-\sigma}\varphi_\alpha
\;,
\eea
respectively.
The modified scalar currents \eqq{msc}
also take a simple form
\bea
{\cal P}_\mu^\alpha = e^{-\sigma}\,(\partial_\mu \varphi^\alpha - g A_\mu^\alpha)
={\cal Q}_\mu^{1\alpha}
\;,\qquad
{\cal P}_\mu^{1} = \partial_\mu \sigma
\ ,
\label{PQcov}
\eea
and the  modified integrability condition \eqq{dp1} and the curvature \eqq{curv1} read
\bea
&& \cD_{[\mu} \cP_{\nu]}^\alpha = -\ft12 g e^{-\sigma} {\cal G}_{\mu\nu}^\alpha \ ,
\nn\w2
&&
\cQ_{\mu\nu}^{1\alpha} = 2\cP_{[\mu}^{\alpha}\,\partial^{\vphantom{\alpha}}_{\nu]} \sigma\,
-g  e^{-\sigma}  {\cal G}_{\mu\nu}^\alpha\,  \ ,
\label{4}
\eea
The above formulae exhibit intricate couplings of vector and tensor fields, and the nature of the shift
symmetries that have been gauged.


\section{Conclusions and discussion }
\label{sec:conc}

In this paper we have determined the possible gaugings in magical supergravities in $6D$, which are supergravities with $8$ real supersymmetries coupled to a fixed number of vector and tensor multiplets, and arbitrary number of hypermultiplets. We have employed the embedding tensor formalism which determines in a systematic fashion the appropriate combination of vector fields that participate in the gauging process. It turns out that the allowed gauge group is uniquely determined in each case and the underlying Lie algebra, displayed in \eqq{algebra}, is nilpotent generated by  $(n_T-1)$ Abelian translations with $(n_T-2)$ central charges. Due to these central charges, the translation generators can not lie strictly in the isometry algebra  $\mathfrak{so}(n_T,1)$ acting on the scalar fields of the tensor multiplets.  The central charges do not act on the vector/tensor sector , but may act nontrivially in the hypersector. We analysed the possible embeddings of the nilpotent gauge group into the isometry groups of the quaternionic K\"ahler manifolds of the hyperscalars. Since  R-symmetry $Sp(1)_R$ is part of the isometry group of the hyperscalars, the embedding of the gauge group  into the isometries on the quaternionic K\"ahler manifold determines whether $U(1)_R$ subgroup of $Sp(1)_R$ can be gauged such that it acts nontrivially on the fermions. In absence of hypermultiplets, the $R$-symmetry acts exclusively on the fermions and one can use a linear combination of the Abelian gauge fields to gauge $U(1)_R$ such that it acts nontrivially on the fermions. It will be interesting to investigate if and how the gaugings obtained by dimensional reduction from these theories fit into existing classifications.

Despite the simultaneous appearance of both Chern-Simons modified $3$-form field strengths as well as
generalized Chern-Simons terms, the gauged magical supergravity theories we have presented are truly gauge invariant.
While arbitrary number of hypermultiplets are allowed, the special number of vector and tensor multiplets is crucial for this invariance.  Indeed, coupling of any additional  vector (and/or tensor) multiplets would impose stringent constraints on the Chern-Simons coupling of the vectors to tensors. Existence and construction of theories satisfying these constraints that can be  interpreted as extensions of  magical supergravities remains to be investigated.  The failure to satisfy these constraints would give rise to classical anomalies which then should satisfy the Wess-Zumino consistency conditions.

Turning to the magical gauged  $6D$ supergravities we have constructed here,  while truly gauge invariant,  they may still have gravitational, gauge and mixed anomalies at the quantum level,  owing to the presence of chiral fermions and self-dual $2$-form potentials.  As is well known, the gravitational anomalies are encoded in an $8$-form anomaly polynomial which, in general, contains terms of the form $({\rm tr} R^4)$ and $({\rm tr} R^2)^2$. The first kind of terms must necessarily be absent for anomaly freedom. In presence of $n_V$ vector multiplets and $n_H$ hypermultiplets, it is well known that this imposes the condition $n_H=273 + n_V-29n_T$. From magical supergravities, this condition is satisfied with multiplicities $(n_T, n_V, n_H)$ given by $(9,16,28)\ ,(5,8,136)\ ,(3,4,190)$ and $(2,2,217)$, respectively \cite{Andrianopoli:2004xu}. Once the condition for the  absence of the $(\tr R^4)$ terms is satisfied, the total gravitational anomaly polynomial (in conventions described in \cite{Salam:1989fm}) becomes $\Omega_8 = \frac{1}{128} (n_T-9) \left(\tr\,R^2\right)^2$, with $\tr\,R^2 \equiv \tr\,R \wedge R$.
The full gravitational anomaly vanishes identically for the octonionic magical supergravity, with $(n_T, n_V, n_H)$ multiplicities given by $(9,16,28)$. However, in presence of gaugings, there will still be
gauge and mixed anomalies. In the gauged magical supergravities with $n_T=2,3,5$, the purely gravitational anomaly will be present as well. The determination of the full set of anomalies and the possible elimination by suitable Green-Schwarz-Sagnotti type mechanism~\cite{Green:1984sg,Sagnotti:1992qw} in gauged magical supergravities is beyond the scope of this paper, and will be treated elsewhere. A detailed analysis is expected to contain elements similar to those encountered in \cite{DeRydt:2008hw} in their treatment of anomalies in gauged $N=1$ supergravities in $4D$ in which the embedding tensor plays a key role as well.

Since their discovery higher dimensional and/or stringy origins of magical supergravity theories , which are invariant under  8 real supersymmetries,  have been of great interest. Largest magical supergravity  defined by the exceptional Jordan algebra $J_3^{\mathbb{O}}$ has groups of the $E$ series as its U-duality group in $5,4$ and $3$ dimensions just like the maximal supergravity with 32 supersymmetries , but are of different real forms. 
The authors of  \cite{Gunaydin:1986fg} 
posed the question whether  the  exceptional $4D$ Maxwell-Einstein supergravity can arise as as the low energy effective theory of type II superstring theory compactified over some exceptional CalabiYau manifold.  
 In the mathematics literature this was posed as the question whether  the scalar  manifold  $\frac{E_{7(-25)}}{E_6\times U(1)}$  of  the exceptional $4D$ Maxwell-Einstein supergravity theory
 could arise as moduli space of deformations of Hodge structures of a Calabi-Yau manifold \cite{MR1258484}. 
 If elliptically fibered, F-theory on such  a Calabi-Yau threefold would  then lead to  the largest $6D$ magical supergravity as its low energy effective theory.  

The maximal supergravity and the largest magical supergravity defined by the octonionic Jordan algebra $J_3^{\mathbb{O}}$ have a common sector which is the magical supergravity theory
defined by the quaternionic Jordan algebra $J_3^{\mathbb{H}}$. 
As was pointed out in \cite{talkparis,Gunaydin:2009pk}  the low energy effective theory of one of the dual pairs of compactifications of IIB superstring to $4D$  studied by Sen and Vafa \cite{Sen:1995ff}  is  precisely the magical $N=2$, $4D$  Maxwell-Einstein supergravity theory defined by the quaternionic Jordan algebra $J_3^{\mathbb{H}}$ without any hypermultiplets\footnote{This follows from the fact that the bosonic sector of the $N=2$ supersymmetric compactification , in question, coincides with that of $N=6$ supergravity. Unique $4D$ Maxwell-Einstein supergravity theory with that property is the quaternionic magical theory \cite{Gunaydin:1983rk}.}. Since the construction of dual pairs in Sen and Vafa's work uses orbifolding on $T^4\times S^1\times S^1$, one can use their methods to construct the $6D$  quaternionic magical theory from $IIB$ superstring directly. Whether one can obtain the gauged quaternionic magical theory constructed above by turning on fluxes is an interesting open problem. The complex magical supergravity defined by $J_3^{\mathbb{C}}$ can be obtained  by truncation of the quaternionic theory to a subsector singlet under a certain $U(1)$ subgroup. In \cite{Dolivet:2007sz} some hypermultiplet-free $N=2$, $4D$  string models based on asymmetric orbifolds  with world-sheet superconformal symmetry using $2D$  fermionic construction were given. Two of these models  correspond to the magical supergravity theories in $4D$ defined by  the complex Jordan algebra $J_3^{\mathbb{C}}$ with the moduli space
\eq \mathcal{M}_4= \frac{SU(3,3)}{SU(3) \times SU(3) \times U(1)} \en ,
and the quaternionic $J_3^{\mathbb{H}}$ theory with the $4D$ moduli space
\eq \mathcal{M}_4= \frac{SO^*(12)}{U(6)} \en
Direct orbifold construction of the exceptional supergravity theory from superstring theory without hypermultiplets has so far proven elusive. In \cite{talkparis,Gunaydin:2009pk} it was argued that an exceptional self-mirror Calabi-Yau 3-fold must exist such that type II superstring theory compactified on it leads to the exceptional supergravity coupled to hypermultiplets parametrizing the quaternionic symmetric space $E_{8(-24)}/E_7 \times SU(2)$. This was based on
the observation  by two of the authors (MG and ES)  that there exists a six dimensional $(1,0)$ supergravity theory, which is free from gravitational anomalies, with 16 vector multiplets, 9 tensor multiplets and 28 hypermultiplets, parametrizing  the exceptional quaternionic symmetric space $E_{8(-24)}/E_7 \times SU(2)$, which reduces to
the $4D$ supergravity with  scalar manifold
\eq \mathcal{M}_V \times \mathcal{M}_H = \frac{E_{7(-25)}}{E_6\times U(1)} \times \frac{E_{8(-24)}}{E_7\times SU(2)}  \en
, and the fact that the moduli space of the FHSV model \cite{Ferrara:1995yx} is a subspace of this doubly exceptional moduli space. The authors of  \cite{Bianchi:2007va} reconsidered the string derivation of  FHSV model  over the Enriques Calabi-Yau manifold, which corresponds to a $6D$, $(1,0)$ supergravity theory with $n_T=9$, $n_H=12$ and $n_V=0$, and argued that the octonionic magical  theory defined by  $J_3^\mathbb{O}$ admits a string interpretation closely related to the Enriques model and  16 Abelian vectors of  the octonionic magical  supergravity theory in  $6D$  is related to the rank of Type I and heterotic strings.
In mathematics literature, Todorov \cite{todorov-2008} gave a construction of a Calabi-Yau 3-fold such that
 Type IIB  superstring theory compactified over it  leads to the magical  $4D$ Maxwell-Einstein supergravity theory defined by the complex Jordan algebra $J_3^{\mathbb{C}}$
 coupled to $ ( h^{(1,1)}+1 ) =30 $ hypermultiplets. Whether his construction can be extended to obtain a Calabi-Yau 3-fold that would lead to the $4D$ exceptional Maxwell-Einstein supergravity theory defined by $J_3^{\mathbb{O}}$ coupled to hypermultiplets is an open problem.  As stated above if  such an exceptional Calabi-Yau 3-fold exists and is elliptically fibered, then F-theory compactified over it is expected to be described by the $6D$ octonionic magical supergravity theory coupled to hypermultiplets.

\subsection*{Acknowledgments}

We wish to thank E. Bergshoeff for useful discussions. We thank each other's home institutions
for hospitality during this work. The work of M.G. was supported in part by the National
Science Foundation under grant numbered  PHY-0855356. Any opinions,
findings and conclusions or recommendations expressed in this
material are those of the authors and do not necessarily reflect the
views of the National Science Foundation.
The work of H.S.\ is supported in part by the Agence Nationale de la Recherche (ANR).
The research of E.S.\ is supported in part by NSF grants PHY-0555575 and PHY-0906222.

\newpage

\providecommand{\href}[2]{#2}\begingroup\raggedright\endgroup
\end{document}